\documentclass[12pt]{article}
\usepackage{setspace}
\usepackage{amsmath, amssymb}
\usepackage{graphicx}
\usepackage{upgreek}
\usepackage{natbib}
\usepackage{hyperref}
\usepackage{authblk}
\usepackage{tikz}
\usetikzlibrary{positioning}

\newtheorem{prop}{Proposition}
\newtheorem{defn}{Definition}

\usepackage{algorithm}
\usepackage{algorithmicx}
\usepackage{algpseudocode}

\title{SoftBART-CORM Models for Density Regression}
\author[1]{J. E. Griffin}
\author[2]{M. Kalli}
\affil[1]{University College London}
\affil[2]{King's College London}
\date{}

\begin{document}

\maketitle

\begin{abstract}

Flexible modelling of the conditional distribution of a response given predictors is important in many applications where nonlinear effects, heteroskedasticity, and multimodality may be present. We introduce the SoftBART-CoRM, a predictor-dependent infinite mixture model that combines the interpretable component structure of Compound Random Measures (CoRMs) with SoftBART priors on both mixture weights and component-specific parameters. The model flexibly captures complex distributional features, while global--local shrinkage and predictor-selection priors mitigate overfitting and identify the covariates driving component allocation and kernel parameters. Posterior inference is performed using an efficient Markov chain Monte Carlo algorithm that leverages existing \texttt{R} implementations of SoftBART. Through simulation studies and real-data applications, we show that the proposed approach can adaptively shrink towards simpler mixture representations, provide interpretable inference on the predictors governing different aspects of the conditional distribution, and achieve superior out-of-sample predictive performance relative to benchmark models.


\end{abstract}

\section{Introduction}

Regression methods have traditionally focused on modelling the conditional mean $\mathbb{E}(y\mid \boldsymbol{x})$ of a response variable $y$ given predictors $\boldsymbol{x}\in\mathbb{R}^p$ to  provide a parsimonious description of their relationship. However, in many applications (from environmental science to economics and social sciences), scientific interest extends beyond the mean to the entire conditional distribution $F(y\mid\boldsymbol{x})$. Predictors can affect not only location, but also variability, skewness, tail behaviour, or even induce multimodality, so that the shape of the response distribution changes across the predictor space. This has motivated the development of distributional and density regression methods whose objective is to estimate the conditional distribution of the response.

Several frameworks have been proposed. Generalised additive models for location, scale and shape (GAMLSS) \citep{Rigby2005} and related Bayesian distributional regression models \citep{Klein2015,Umlauf2018} allow multiple shape-defining parameters of a chosen parametric family to be functions of predictors. Quantile regression \citep{Koenker1978} instead models conditional quantiles directly, while conditional transformation models \citep{chern2013,Hothorn2014} represent the conditional distribution through a predictor-dependent transformation of a reference distribution. These approaches substantially extend classical mean regression, but their flexibility ultimately remains tied to a pre-specified distributional family, a finite collection of quantiles, or a particular transformation representation. Consequently, they may struggle to capture highly complex conditional distributions characterised by multimodality, heterogeneous tail behaviour, local irregularities, or abrupt changes across the predictor space. For a review of these methods see \cite{Klein2024}.

Bayesian nonparametric mixture models provide a particularly attractive alternative because they place minimal restrictions on the shape of the response distribution and allow the complexity of the model to adapt to the data. Following the seminal work of \citet{MacEachern1999} on dependent Dirichlet processes (DDP), a large literature has developed around infinite predictor-dependent mixture models of the form 
 \begin{equation} 
 f(y\mid\boldsymbol{x}) = \sum_{k=1}^{\infty} w_k(\boldsymbol{x}) \,K\!\left(y;\theta_k(\boldsymbol{x})\right), 
 \label{eq:general-mixture} \end{equation} 
where $K(y; \theta)$ is a kernel with parameters $\theta$, and both the mixing weights, $w_k(\boldsymbol{x})$, and kernel parameters, $\theta_k(\boldsymbol{x})$, are functions of  the predictors.  We focus on the choice $\theta(\boldsymbol{x}) = \left(\mu(\boldsymbol{x}), \sigma^2(\boldsymbol{x})\right)$ and a normal kernel distribution, $K(y; \theta(\boldsymbol{x})) = \mbox{N}\left(y; \mu(\boldsymbol{x}), \sigma^2(\boldsymbol{x})\right)$.
Such models can account for changing multimodality, skewness, heteroscedasticity and tail behaviour across the predictor space
 while retaining the adaptivity properties of Bayesian nonparametric mixtures. Theoretical support for these models is substantial. 
Dirichlet process mixtures have large support on broad classes of distributions \citep{Lo1984,Ghosal1999}, while posterior consistency has been established under suitable regularity conditions \citep{Ghosal2001,Ghosal2007}. Related support and consistency results are now available for broad classes of predictor-dependent mixture models \citep{Tokdar2007,BaJaQu12,Pati2013,Norets2014}. Recent overviews are provided by \citet{quin22} and \citet{WaIn25}. 

  A challenge in infinite predictor-dependent mixture models is building sufficient flexibility to capture complex  relationships between the predictors and the response distribution,
 while remaining computationally tractable and avoiding overfitting in moderate- and high-dimensional settings. Much of the existing literature has focussed on 
 allowing either the weights or the kernel parameters to depend on linear functions of the predictors, but not both. For example, the single-$p$ DDP models of \citet{Mac00} and the ANOVA-DDP model of \citet{iorio04} introduce predictor dependence through the kernel parameters while keeping the weights fixed. Conversely, predictor-dependent stick-breaking processes and related constructions often introduce dependence solely through the weights \citep{griste06,dunson08,dunson11,GriLei17}. Although these restrictions simplify posterior computation and can yield consistent estimation, empirical evidence suggests that they can severely restrict the model's ability to capture complex predictor-dependent distributional features, often resulting in poor finite-sample performance \citep{BaJaQu12,WaIn25}. 

We address this  challenge  using separate non-linear functions for both the weights ($w_k(\boldsymbol{x})$) and the kernel parameters ($\mu_k(\boldsymbol{x})$ and $\sigma_k^2(\boldsymbol{x})$) in each component in (\ref{eq:general-mixture}). This leads to a very general normalized Compound Random Measure (NCoRM) regression model  \citep{GriLei17, GriLei18}, where $\mu_k(\boldsymbol{x})$ and $\sigma_k^2(\boldsymbol{x})$ can respond to predictors in fundamentally different ways in each component of the mixture. The non-linear functions are modelled using a SoftBART prior \citep{LiYa18}, which develops the DART version \citep{Lin18} of the BART prior  \citep{ChiGeoMcc10}, to allow continuous regression functions from a Bayesian tree ensemble and perform automatic variable selection.
The BART prior and its generalisations can model functions with complex nonlinear effects and higher-order interactions in moderate-dimensional settings whilst avoiding overfitting 
using  a carefully designed prior distribution   \citep[see][for a comprehensive review]{hill20}.  \citet{RoVDP20}  and \citet{LiYa18} study the theoretical properties of BART-type models and show that they can achieve optimal minimax rates for recovering smooth surfaces and automatically  adapt  to unknown levels of smoothness. Modelling the weights and the kernel parameters using separate non-linear functions with SoftBART priors leads to a highly-parametrized model. Global-local shrinkage priors are used to control flexibility and avoid overfitting.  To proceed to Bayesian inference we use a Markov chain Monte Carlo (MCMC) algorithm that leverages existing highly optimised SoftBART software and standard data augmentation approaches. In addition, the simple structure in \eqref{eq:general-mixture} and the in-build automatic variable selection of SoftBART provides an interpretable approach. We call our modelling framework SoftBART-CoRM.

The paper is organised as follows, Section 2 provides an overview of BART models and their extensions while Section 3 introduces the SoftBART-CoRM framework, describes the prior specification, and discusses its flexibility, interpretability, and shrinkage properties. Section 4 provides an overview of the MCMC algorithm with full details included in Appendix~(\ref{MCMC_full}) of the supplementary material. Section 5 explains how the model can be interpreted using standard label-switching algorithms, variable selection and partial dependency plots. Section 6 applies the SoftBART-CoRM model  to both simulated and real data, and compares predictive performance to some standard benchmarks. Section 7 provides concluding remarks and future research directions.

\section{Bayesian Additive Regression Trees models}

Bayesian Additive Regression Trees (BART) models \citep{ChiGeoMcc10} are highly scaleable and flexible models of the conditional mean $\eta(\boldsymbol{x}) = \mathbb{E}(y|\boldsymbol{x})$ which  capture non-linear relationships and multi-way interactions between variables whilst avoiding overfitting. The BART model assumes normal errors, 
\begin{equation}
y = \eta(\boldsymbol{x}) + \epsilon, \qquad \epsilon\sim \mbox{N}(0, \upsigma^2).
\label{basic-nonlinear}
\end{equation}
The regression function $\eta(\boldsymbol{x})$ is modelled by an ensemble of $R$ trees $\mathcal{T} = (T_1, \dots, T_R)$ with leaf parameters $\mathcal{A} = (\boldsymbol\alpha_1, \dots, \boldsymbol\alpha_R), $ 
\begin{equation}
\eta(\boldsymbol{x}) =  \sum_{r = 1}^R g(\boldsymbol{x}; T_r, \boldsymbol\alpha_r),\label{plainbart}
\end{equation}
where the function $g(\boldsymbol{x}; T_r, \alpha_r)$ is the value given to $\boldsymbol{x}$ by the tree $T_r$ with leaf parameters $\boldsymbol\alpha_r$.  
The $r$-th tree $T_r$ is defined by 
a sequence of $L_r-1$
axis-aligned splits of the form  $\{x_{j_{r, \varsigma}} \leq C_{r, \varsigma}\}$ and $\{x_{j_{r, \varsigma}} > C_{r, \varsigma}\}$ for $\varsigma = 1, \dots, L_r-1$
 where $x_{j_{r, \varsigma}}$ and $C_{r, \varsigma}$ are the splitting variable and threshold of the $\varsigma$-th split in the $r$th tree. A parameter is associated with each leaf of $T_r$, and all leaf parameters are collected as $\boldsymbol\alpha_r = (\alpha_{r, 1}, \dots, \alpha_{r, L_r})$.

 \citet{ChiGeoMcc10} define default priors for the tree and leaf parameters to allow the general use of their method.
 The response $y$ is  normalised to be between $-0.5$ and $0.5$ and each predictor $x_i$ to be between $0$ and $1$. The pairs $(T_1, \boldsymbol\alpha_1), \dots (T_R, \boldsymbol\alpha_R)$ are given  independent priors with the form $\pi(T_r, \boldsymbol\alpha_r) = \pi(\boldsymbol\alpha_r\vert T_r)\,\pi(T_r)$. The prior for $T_r$ 
is defined sequentially with a split at the $m$-th level of the tree occurring with probability $\pi_{m}$. At each split in the tree, the splitting variable is chosen at random and the threshold is chosen uniformly from the observed predictor values for the splitting variable. Conditional on the tree, the prior for 
the $l$-th leaf parameter of the $r$-th tree is $\alpha_{r, l}\sim \mbox{N}(0, \kappa^2)$. This implies  $\eta(\boldsymbol{x}) \sim \mbox{N}(0, R\,\kappa^2)$
and so $\kappa^2$ controls the prior range of the conditional mean over $\boldsymbol{x}$. \citet{ChiGeoMcc10}  suggest a default value for $\kappa^2$, which exploits 
the normalisation of the response, and \citep{Lin18} suggests a half-Cauchy prior distribution.

The BART model in (\ref{basic-nonlinear}) assumes normal errors with a constant $\upsigma^2$ across all values of $\boldsymbol{x}$ which can be restrictive. In many applications (medical, social, environmental and financial) assuming the same level of noise across the entire dataset is detrimental for both predictive performance and uncertainty quantification. This limitation  led to a range of models which allow weaker assumptions. Parametric extensions include heteroscedastic (HBART) \citep{PrChGeMc20} and skew (SBART) \citep{UmLiSiBa23}. 
To avoid parametric assumptions for the error distribution, \cite{GeLaLoMcSp19} use a Dirichlet process (DP) mixture of normals with an unknown constant $\upsigma^2$, while \cite{SpLoMaLaMc23} extend this approach to a heteroscedastic error model. Some extensions move away from only modelling $\mathbb{E}(y|\boldsymbol{x})$ using BART. The HBART introduces a second BART for the conditional variance $\mbox{Var}(y|\boldsymbol{x})$, while \citet{KiWaHaPe16}  and  \citet{LiLuSVCh25} model the conditional density $f(y|\boldsymbol{x})$ using quantile regression where the quantile functions are BARTs. This has led to the use of BARTs in conditional transformation models where the predictor transformation function is given a heteroscedastic BART prior \citep{OrMuLiVo21}. An alternative model is that of \citet{LiLiMu23} where $f(y|\boldsymbol{x})$ is modelled by tilting a parametric model (a normal linear regression or a normal model) by a probit or logit transform of a BART-type model.
Although these models provide more flexibility, they still assume either parametric distributions for the errors or restricted forms of nonparametric modelling which can lead to poor estimation.

The SoftBART model \citep{LiYa18} extends BART to smooth, continuous functions that better capture physical processes and incorporates automatic variable selection through the DART prior of \citep{Lin18}.  To generate the continuous regression functions SoftBART replaces the hard axis-aligned splits  $\{x_{j_{r, \varsigma}} \leq C_{r, \varsigma}\}$ with a soft version which assigns an observation to the left tree branch with probability $\psi((x_{j_{r, \varsigma}} - C_{r, \varsigma}) / \tau_b)$, where $\tau_b$ is a bandwidth parameter.  The soft splits will behave increasingly like the hard splits  as $\tau_b$ becomes close to $0$. We follow \citet{LiYa18} and choose the logistic function $\psi(x) = 1 / (1 + \exp\{-x\})$ and $\tau_b\sim \mbox{Exp}(r)$ where $\mathbb{E}(\tau_b) = r$ with $r$ chosen to be $0.1$. The DART prior of \citet{Lin18} generalizes the original BART method by assuming that the $j$-th predictor is chosen with probability $s_j$ at any split in any tree. These probabilities are often referred to as predictor importance parameters and  are given the prior $\boldsymbol{s} = (s_1,\dots, s_p)\sim \mbox{Dirichlet}(a/p, \dots, a/p)$. A larger value of $s_j$ implies that  the $j$-th variable is more useful for explaining the variation in $\eta(\boldsymbol{x})$ and so $\boldsymbol{s}$  can be used to perform automatic variable selection. We will write the SoftBART prior with parameters $\kappa$ and $\boldsymbol{s}$ as $\mbox{SoftBART}(\kappa, \boldsymbol{s})$.

\section{The SoftBART-CoRM model}

Our flexible Bayesian nonparametric framework for density regression models the conditional density of $y$ given predictors $\boldsymbol{x}$ using a predictor-dependent infinite mixture of the form, \begin{equation} 
f(y\mid\boldsymbol{x}) = \sum_{k=1}^{\infty} w_k(\boldsymbol{x}) \, \mathrm{N}\!\left( y; \mu_k(\boldsymbol{x}), \sigma_k^2(\boldsymbol{x}) \right), 
\label{eq:normal-mixture} 
\end{equation} 
where both the mixture weights $w_k(\boldsymbol{x})$ and the kernel parameters $\mu(\boldsymbol{x})$ and $\sigma^2(\boldsymbol{x})$ are flexible functions. 
We refer to $w_k(\boldsymbol{x})$, $\mu(\boldsymbol{x})$ and $\sigma^2(\boldsymbol{x})$ as elements of the mixture model.
This very general specification allows us to capture complex nonlinearities and interactions in any element of the mixture model without requiring their explicit specification. 
The resulting model is typically more parsimonious and more interpretable than models with either constant weights or constant kernel parameters. For example, if component means depend on predictors while kernel parameters are held constant,  additional mixture components  must be introduced to capture this covariate-dependent variation, thereby increasing model complexity.
As is standard in BART-type models, we assume that the response has been standardised so that $y_i\in(-0.5,0.5)$ and that each predictor  lies in $(0,1)$. 
 Retaining this standardisation has two important advantages. First, it allows a common default prior specification across different data sets. Second, it places all predictors on a comparable scale, which is particularly important for variable selection.
 
  We define the SoftBART-CoRM as follows:
\begin{defn}
Let  $y\in \mathbb{R}$ be a response variable and $\boldsymbol{x}\in \mathbb{R}^p$ be a vector of predictors, a SoftBART-CORM model assumes that 
\begin{equation}
f(y\mid \boldsymbol{x}) = \sum_{k=1}^{\infty} \frac{J_k\,\exp\{\lambda_k(\boldsymbol{x})\}}{\sum_{k'=1}^{\infty} J_{k'}\, \exp\{\lambda_{k'}(\boldsymbol{x})\}}
 \,\mbox{N}\left(y\left\vert \mu_k + \delta_k(\boldsymbol{x}), \sigma^2_k \, \exp\{\phi_k(\boldsymbol{x})\}\right.\right)
 \label{def-BART-CORM}
 \end{equation}
with the prior
\[
 \lambda_k(\boldsymbol{x})\stackrel{i.i.d.}{\sim} \mbox{SoftBART}(\kappa^{\lambda}_k, \boldsymbol{s}^{\lambda}),\qquad k = 1, 2, \dots
 \] 
 \[
\mu_k\stackrel{i.i.d.}{\sim} \mbox{N}(0, 1), \quad \delta_k(\boldsymbol{x})\stackrel{i.i.d.}{\sim}\mbox{SoftBART}(\kappa^{\delta}_k, \boldsymbol{s}^{\delta}), \qquad k = 1,2, \dots
 \]
 \[
\sigma^2_k\stackrel{i.i.d.}{\sim} \mbox{Ga}(a, b),\quad  \phi_k(\boldsymbol{x})\stackrel{i.i.d.}{\sim} \mbox{SoftBART}(\kappa^{\phi}_k, \boldsymbol{s}^{\phi})
\]
and $J_1, J_2,  \dots$ are the jumps of a completely random measure (CRM) with L\'evy intensity $\nu(J).$   
\end{defn} 
The continuity of the distribution function $F(y\mid \boldsymbol{x})$ in $\boldsymbol{x}$ clearly follows from  
the continuity of the functions  $\delta_r(\boldsymbol{x})$, $\phi_r(\boldsymbol{x})$, and $\lambda_r(\boldsymbol{x})$ generated by a SoftBART prior.
The model can also be written as an NCoRM mixture model \citep{GriLei17} by writing
\[
p(y\mid \boldsymbol{x}) = \int 
 \mbox{N}(y\mid \mu, \sigma^2 )\, dF_{\boldsymbol{x}}(\mu, \sigma^2)
\]
where
\[
F_{\boldsymbol{x}}(\mu, \sigma^2) 
= \sum_{k=1}^{\infty} \frac{J_k\,\exp\{\lambda_k(\boldsymbol{x})\}}{\sum_{k'=1}^{\infty} J_{k'}\, \exp\{\lambda_{k'}(\boldsymbol{x})\}} 
\delta_{\mu = \mu_k +  \delta_k(\boldsymbol{x}), \sigma^2 = \sigma^2_k \, \exp\{\phi_k(\boldsymbol{x})\}}.
\]
The existence of the model follows from applying the integrability conditions for compound random measures provided by  \cite{RPLe18}.
\begin{prop} 
The SoftBART-CoRM provides well-defined marginal processes for all $\boldsymbol{x}$.
\end{prop}
 
 The SoftBART prior has mean zero and so each element of the mixture in the SoftBART-CoRM is given an offset,
 $\mu_k(\boldsymbol{x}) = \mu_k + \delta_k(\boldsymbol{x})$ and $\sigma^2(\boldsymbol{x}) = \sigma^2_k \exp\{\phi_k(\boldsymbol{x})\}$, which yields a highly flexible modelling framework where each element of the model varies around the corresponding constant offset. 
 The use of $\phi_k(\boldsymbol{x})$ as a regression function on the log-variance scale is similar in spirit to \citet{OrMuLiVo21}, while differing substantially from heteroscedastic BART models such as \citet{PrChGeMc20}, which specify conjugate variance priors at the terminal-node level.
 Similarly, the weight can be written as $w_k(\boldsymbol{x}) = \frac{J_k(\boldsymbol{x})}{\sum_{k'=1}^{\infty} J_{k'}(\boldsymbol{x})}$ where $J_k(\boldsymbol{x}) = J_k \exp\{\lambda_k(\boldsymbol{x})\}$. The use of a completely random measure for $J_1, J_2, \dots$ encourages unnecessary mixture components to receive negligible weight in the same spirit as standard Bayesian nonparametric mixture models. 
 The standardisation of the response allows us to define a default prior for the mean offset $\mu_k$ and the variance offset $\sigma^2_k$. 
 A wide range of simpler models can be defined through parameter restrictions. A SoftBART model with Dirichlet process errors (similar to the model in \citet{GeLaLoMcSp19}) arises by setting $\delta_k(\boldsymbol{x})=\delta(\boldsymbol{x})$, $\lambda_k(\boldsymbol{x}) = 0$, $\phi_k(\boldsymbol{x}) = 0$ for $k=1,2,\ldots$, and $\nu$ to be the L\'evy intensity of a gamma process. A heteroscedastic counterpart, in the spirit of \citet{SpLoMaLaMc23}, is obtained by  instead imposing $\phi_k(\boldsymbol{x})=\phi(\boldsymbol{x}),$ for $k=1,2,\ldots$. 
 A weights-only variation arises from
setting  $ \delta_k(\boldsymbol{x})=0$ and $\phi_k(\boldsymbol{x})=0, $ for $k=1,2,\ldots$. A model with constant weights in the spirit of ANOVA-DDP  \citep{iorio04} 
  is defined by setting $\lambda_k(\boldsymbol{x}) = 0$. Further simpler models can be defined by restricting 
   other combinations of functions. For example, a model with predictor-dependent weights and means is defined by setting $\phi_k(\boldsymbol{x}) = 0$.
 
A key advantage of the SoftBART-CoRM framework is its use of a CoRM prior rather than the stick-breaking constructions typically adopted in dependent Dirichlet process (DDP) mixtures. Stick-breaking priors impose an artificial ordering on the mixture components. Consequently, the effect of the predictors on any weight will propagate through to the subsequent weights, making component interpretation difficult and complicating the characterisation of dependence across predictor values. In contrast to stick-breaking constructions, CoRM priors yield exchangeable weights whose interpretation is preserved across the collection of random measures. Dependence is introduced through latent score functions, allowing flexible borrowing of information while maintaining a coherent component structure.

The use of SoftBART priors provides two key advantages. First, standard SoftBART variable selection tools can be used to quantify the importance of each predictor for the component means, variances, and mixing weights. Second, the piecewise-constant leaf parameters induce natural regularisation of the component functions in regions of predictor space where a particular component is poorly represented. This helps avoid unstable estimates of component means, variances, and weights, resulting in more robust conditional density estimation. Unlike standard regression extrapolation, this issue can arise even in regions containing observations, since only a small subset of those observations may be associated with a given mixture component.

The SoftBART-CoRM prior is completed by specifying hyperpriors for the scale parameters $\kappa^{\lambda}_k$, $\kappa^{\delta}_k$ and $\kappa^{\phi}_k$, and the predictor importance parameters $\boldsymbol{s}^{\lambda}$, $\boldsymbol{s}^{\delta}$ and $\boldsymbol{s}^{\phi}$. The specification is designed to control model complexity while retaining sufficient flexibility to capture important predictor effects.
 As previously discussed setting some (or all) of the functions $\delta_k(\boldsymbol{x})$, $\phi_k(\boldsymbol{x})$, and $\lambda_k(\boldsymbol{x})$ to zero defines a simpler model where some element of the mixture model does not depend on the predictors (only the offsets). The prior variance of these functions is
  $\kappa_k^{\zeta}\,R$, where $R$ is the number of trees and $\kappa^{\zeta} \,\,\mbox{for} \,\,\zeta \in \{\delta,\phi,\lambda\}$ is a leaf-scale parameter. 
 This suggests using global-local shrinkage priors on the leaf-scale parameters to encourage these functions to be shrunk  towards zero. We use a half-Cauchy distribution with parameter $\rho$, denoted as
 $x\sim\mathrm{HC}(\rho)$ with density
  \[ 
  p(x)=\frac{2}{\pi\rho} \left(1+\frac{x^2}{\rho^2}\right)^{-1}, \qquad x>0.
  \] 
The assigned priors to the leaf parameters are,  for each $\zeta\in\{\delta,\phi,\lambda\}$,
  \[
   \kappa_k^\zeta \stackrel{i.i.d.}{\sim} \mathrm{HC}(\rho^\zeta), \qquad k=1,2,\ldots.
   \] 
Each element of the model operates on a different scale, and interacts with the corresponding offset, making it difficult to determine suitable values of the global scale parameters $\rho^{\delta}$, $\rho^{\phi}$ and $\rho^{\lambda}$ a priori. We therefore extend the model hierarchy by introducing half-Cauchy hyperpriors on the global scale parameters
\[ \rho^\zeta \stackrel{i.i.d.}{\sim}\mathrm{HC}\!\left( \frac{1}{4\sqrt{R}} \right), \qquad \zeta\in\{\delta,\phi,\lambda\}. \] 

This specification allows different components to exhibit different levels of predictor dependence (through different $\kappa_k^{\zeta}$ values) while sharing information through the common scale parameter $\rho^\zeta$. Components with little support for predictor effects are therefore free to remain close to their offsets, whereas components requiring substantial variation are not overly constrained. The heavy tails of the half-Cauchy distribution prevent excessive penalisation of large function effects and allow substantial heterogeneity across mixture components. 

Each predictor importance parameter is given a  DART prior \[ \boldsymbol{s}^{\delta}, \boldsymbol{s}^{\phi}, \boldsymbol{s}^{\lambda} \stackrel{\text{i.i.d.}}{\sim} \mathrm{Dir}_p(1/p,\ldots,1/p), \] which encourages sparsity and allows irrelevant predictors to be excluded from the regression functions.  Additionally,  the split probabilities are shared across all mixture components, allowing information about predictor importance to be pooled.

\section{Computational approach}

\cite{GriLei18} developed MCMC methods for NCoRM density regression models that could be adapted to the SoftBART-CoRM framework. Our primary objective, however, is to retain compatibility with existing and computationally efficient BART software, which allow the response, predictors and hyperparameters of a BART regression model to be modified during an MCMC run (such as the R packages \verb+SoftBart+ \citep{Li22} or \verb+dbarts+ \citep{DoChMc26}). We  adopt a finite truncation of the underlying completely random measure, $J_1,J_2,\ldots,J_K,$ with fixed truncation level $K$. Throughout the paper we consider the finite approximation of a gamma process, $ J_k \stackrel{\mathrm{i.i.d.}}{\sim} \mbox{Ga}(M/K,1), \,\, k=1,\ldots,K,$ where $M$ is the mass parameter. This specification leads to a finite mixture model with shrinkage on the component weights and can therefore be viewed as an overfitted mixture model \citep{RoMe11}. The same strategy can be applied to other completely random measures admitting convenient finite-dimensional representations, while more general processes could be accommodated through efficient implementations of the Ferguson--Klass algorithm \citep{BerGri26}.

MCMC methods for mixture models introduce latent variables that link each observation to a mixture component, which we define to be $\boldsymbol{q} = (q_1, \dots, q_n)$ leading to the following representation of the SoftBART-CoRM model:
\begin{align}
y_i \mid \boldsymbol{x}_i,q_i=k
&\sim
\mbox{N}\!\left(
\mu_k+\delta_k(\boldsymbol{x}_i),
\,\sigma_k^2\exp\{\phi_k(\boldsymbol{x}_i)\}
\right), \nonumber\\
p(q_i=k\mid\boldsymbol{x}_i)
&=
\frac{J_k\exp\{\lambda_k(\boldsymbol{x}_i)\}}
{\sum_{r'=1}^{K}
J_{r'}\exp\{\lambda_{r'}(\boldsymbol{x}_i)\}}.
\label{def-BART-CORM-2}
\end{align}
Full details of the MCMC algorithm are given in Appendix~\ref{MCMC_full}. Here we highlight the main computational ideas. The model contains the functions $\delta_k(\boldsymbol{x})$, $\phi_k(\boldsymbol{x})$ and $\lambda_k(\boldsymbol{x})$, together with corresponding offset parameters $\mu_k$, $\sigma_k^2$ and $J_k$. Strong posterior dependence between these quantities can lead to poor Gibbs mixing and so we incorporate interweaving steps \citep{YuMen11}.

A key feature of the proposed sampler is that conditional on the latent allocation variables $\boldsymbol{q}$ all functional updates reduce to  regression problems with normal errors that can be handled using existing BART software.  Conditional on $\boldsymbol q$, the likelihood contribution for $\delta_r(\boldsymbol{x})$ is
a weighted SoftBART regression model with responses $y_i-\mu_r(\boldsymbol{x}_i)$ and observation weights 
$\{\sigma_r^2\exp(\phi_r(\boldsymbol{x}_i))\}^{-1}$, restricted to observations satisfying $q_i=r$.
To update $\phi_r$, conditional on $\boldsymbol{q}$, we re-write the normal kernel part of the SoftBART-CORM in (\ref{def-BART-CORM-2}) as 
\[
\epsilon_i=
\frac{
y_i-\mu_{q_i}-\delta_{q_i}(\boldsymbol{x}_i)
}
{\sigma_{q_i}},
\]
where $\epsilon_i\sim\mbox{N}\!\left(0,\exp\{\phi_{q_i}(\boldsymbol{x}_i)\}\right)$.
The approach of \citet{OCSN07}, which express the distribution of $\log\epsilon_i^2$ as a 10-component mixture model, allows us to define a conditionally normal likelihood. 
The likelihood contribution for $\lambda_r(\boldsymbol{x})$ can be written as
a multinomial logistic regression model. P\'olya--Gamma data augmentation \citep{PSW13} yields a conditionally Gaussian likelihood.

\section{Interpretation and variable selection}
\label{s:interpretation}

The SoftBART-CoRM model expresses the conditional distribution of the response given the predictors as a mixture model, enabling interpretation through component-specific elements of the mixture model, $w_k(\boldsymbol{x})$, $\mu_k(\boldsymbol{x})$ and $\sigma^2_k(\boldsymbol{x})$. Since mixture models are identifiable only up to relabelling of the components, we apply the first iterative version of the Equivalence Class Representative (ECR) algorithm \citep{RodWal14}, implemented in the \verb+label.switching+ R package \citep{Pa16}, to post-process the MCMC output. This allows coherent posterior inference on individual components. We focus on the components with the largest average weights ($\frac{1}{n}\sum_{i=1}^n w_k(\boldsymbol{x}_i)$) and examine how the predictors influence $\mu_k(\boldsymbol{x})$, $\sigma^2_k(\boldsymbol{x})$, and $w_k(\boldsymbol{x})$, respectively. 

We adopt a three-stage strategy to interpret the effects of predictors on each element of the mixture model. First we assess whether there is evidence of predictor-driven variation in each element of the mixture ($\mu_k(\boldsymbol{x})$, $\sigma^2_k(\boldsymbol{x})$, and $w_k(\boldsymbol{x})$).
Second, if there is predictor-driven variation, we identify the predictors that explain it. Finally, we use partial dependence plots (PDP) to visualise the relationship between  the important predictors for each element.

The leaf-scale parameters $\kappa_k^{\delta}$, $\kappa_k^{\phi}$ and $\kappa_k^{\lambda}$ control predictor-driven variation in the component means, variances and weights respectively. 
and we assess their posterior medians to  assess if such variation is present.
Since $\delta_k(\boldsymbol{x})$, $\phi_k(\boldsymbol{x})$, and $\lambda_k(\boldsymbol{x})$ have different scales, there is no universally meaningful threshold for determining whether a value of $\kappa_k$ indicates substantial variation. 
If there is evidence of variation in either $\delta_k(\boldsymbol{x})$, or $\phi_k(\boldsymbol{x})$, or $\lambda_k(\boldsymbol{x})$ or any combination of these for at least one mixture component, we examine the posterior means of the corresponding predictor inclusion probabilities $\boldsymbol{s}^{\delta}$, $\boldsymbol{s}^{\phi}$ or $\boldsymbol{s}^{\lambda}$. These probabilities are interpreted analogously to the DART prior, with larger values indicating greater predictor importance. As in DART, the Dirichlet prior encourages sparsity, often concentrating posterior mass on a relatively small subset of predictors within each element of the mixture. 

We use partial dependence plots  (PDP) \citep{Fri01} to visualise the effects of the predictors on each element of the mixture. PDP's are often used in tree-based models, where the effect of a predictor is measured by averaging the regression function evaluated at values of this predictor and the sampled values of the other predictors.
This provides a useful summary of the effect but can be hard to interpret if the $j$-th predictor is highly correlated with some of the other predictors.
Since we want to understand the effect for the $k$-th component, we adapt the method by averaging only over those observations that are allocated to the $k$-th component using the allocation variables $q_1, \dots, q_n$ in the representation in 
\eqref{def-BART-CORM-2}. Therefore, we average effect  of the $j$-th predictor in the $k$-th component to be
\[
\bar{\eta}_{j, k}(z) =\frac{\sum_{i=1}^n \mbox{I}(s_i=k)\, \eta_k(x_{i, 1}, \dots, x_{i, j-1}, z, x_{i, j + 1}, \dots, x_{i, p})}{\sum_{i=1}^n I(q_i = k)}  
\]
where $\eta_k(\cdot)$ is one of the elements of SoftBART-CoRM model, {\it i.e.} $w_k(\boldsymbol{x})$, $\mu_k(\boldsymbol{x})$, or $ \sigma^2_k(\boldsymbol{x})$. 


\section{Examples}

\subsection{Simulated Examples}\label{sim}

To demonstrate the flexibility of the SoftBART-CoRM framework, we consider three simulated regression examples.
For each example we generate $n=1000$ observations, where the predictors are  $x_{i, j} \stackrel{i.i.d.}{\sim} \mbox{U}(0, 1)$ and 
the response $y_i$ is drawn from a three component mixture of regressions model
\begin{equation}
 p(y_i\vert \boldsymbol{x}_i) =  \sum_{k=1}^3 w_k(\boldsymbol{x}_i) \, \mbox{N}\left(y_i \left\vert \mu_k(\boldsymbol{x}_i), \sigma^2_k(\boldsymbol{x}_i)\right.\right), \qquad i = 1, \dots, n.
\label{simex}\end{equation}
For each example, we fit a SoftBART-CoRM with $K=10$ and run our MCMC algorithm for 17500 iterations using 5000 as burn-in with the remaining samples thinned every $5$th value resulting a final MCMC sample of size 2500. 

The three simulated examples are chosen to demonstrate the adaptive nature of the SoftBART-CoRM framework. In particular, they show that the model can shrink to a parsimonious finite mixture representation when warranted by the data, capture predictor-dependent heteroscedasticity, and simultaneously perform effective variable selection. These examples are listed below:

\begin{enumerate}
\item The response is independent of the univariate predictor ($p=1$), with mixture weights $w_1(x_i)=0.4,\,\, w_2(x_i)=0.4,\,\, w_3(x_i)=0.2, $ means $\mu_1(x_i)=-2,\,\, \mu_2(x_i)=0,\,\, \mu_3(x_i)=2,$ and variances $\sigma_k^2(x_i)=0.5^2,\,\, k=1,2,3.$ As both the mixture weights and kernel parameters are constant across the predictor space, the data-generating mechanism reduces to a finite Gaussian mixture model. This setting is included to evaluate whether the SoftBART-CoRM framework can adaptively shrink to a simpler finite mixture representation when no predictor dependence is present.

\item The response depends on a single predictor ($p=1$) through the mixture weights, component means, and component variances. Specifically, the mixture weights are defined by $w_1(x_i) \propto \exp\{-1+2x_i\}, \,\, w_2(x_i) \propto \exp\{1-2x_i\}, \,\,w_3(x_i) \propto 1,$ with normalisation ensuring that $\sum_{k=1}^{3} w_k(x_i)=1$. The component means are $\mu_1(x_i)=-2+x_i, \,\, \mu_2(x_i)=0, \,\, \mu_3(x_i)=2-x_i,$  and the corresponding variances are $\sigma_1^2(x_i)=0.25^2\exp\{0.6-1.2x_i\}, \,\,\sigma_2^2(x_i)=0.25^2, \,\, \sigma_3^2(x_i)=0.25^2\exp\{-0.6+1.2x_i\}.$ This setting provides a comprehensive assessment of the SoftBART-CoRM framework by requiring the simultaneous recovery of predictor effects on the mixture weights, means, and variances of the component distributions.


\item 
The response depends on  two predictors ($p=2$). The component means depend only on the first predictor, the component variances depend only on the second predictor, and the mixture weights depend on both predictors. The mixture weights are specified as $w_1(\boldsymbol{x}_i) \propto \exp\{-1+2x_{i,1}\}, \,\,w_2(\boldsymbol{x}_i) \propto \exp\{1-2x_{i,2}\}, \,\, w_3(\boldsymbol{x}_i) \propto 1,$ with normalisation such that $\sum_{k=1}^{3} w_k(\boldsymbol{x}_i)=1$. The component means are $\mu_1(\boldsymbol{x}_i)=-2+x_{i,1}, \,\,\mu_2(\boldsymbol{x}_i)=0, \,\, \mu_3(\boldsymbol{x}_i)=2-x_{i,1},$ while the component variances are given by $\sigma_1^2(\boldsymbol{x}_i)=0.25^2\exp\{0.6-1.2x_{i,2}\}, \,\, \sigma_2^2(\boldsymbol{x}_i)=0.25^2, \,\, \sigma_3^2(\boldsymbol{x}_i)=0.25^2\exp\{-0.6+1.2x_{i,2}\}.$  This example assesses the ability of the SoftBART-CoRM framework to perform variable selection across the different model components by identifying which predictors drive the mixture weights, component means, and component variances.

\end{enumerate}

\begin{figure}[h!]
\centering
\includegraphics[trim= 0mm 86mm 0mm 60mm, clip, scale=0.6]{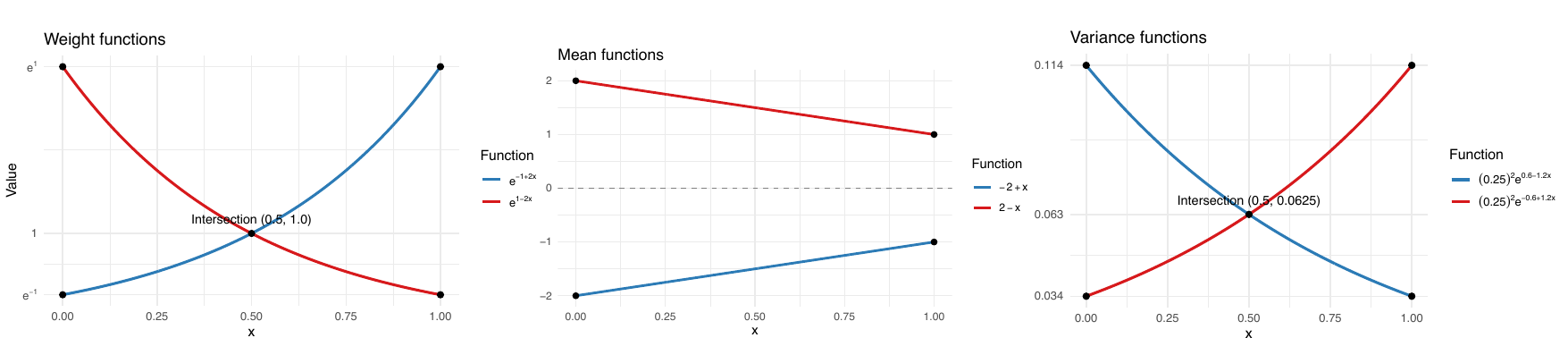}
\caption{Profiles of functions $\gamma_k(\boldsymbol{x}_i)$, $\mu_k(\boldsymbol{x}_i)$, and $\sigma^2_k(\boldsymbol{x}_i)$ chosen in our simulated examples}\label{profs}
\end{figure}
\begin{figure}[h!]
 \centering
  \begin{tabular}{ccc} 
  \multicolumn{3}{c}{\textbf{Example 1}}\\[2mm] 
  \multicolumn{3}{c}{ \begin{tabular}{ccc} Mean & S.D. & Weights\\ \includegraphics[trim=0mm 0mm 0mm 5mm,clip]{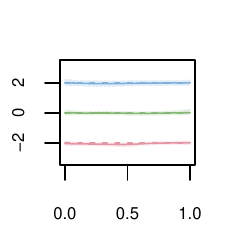} & \includegraphics[trim=0mm 0mm 0mm 5mm,clip]{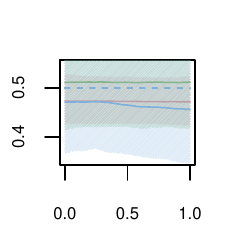} & \includegraphics[trim=0mm 0mm 0mm 5mm,clip]{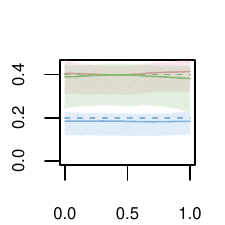} \end{tabular} }\\[5mm] 
  \multicolumn{3}{c}{\textbf{Example 2}}\\[2mm] \multicolumn{3}{c}{ \begin{tabular}{ccc} Mean & S.D. & Weights\\ \includegraphics[trim=0mm 0mm 0mm 5mm,clip]{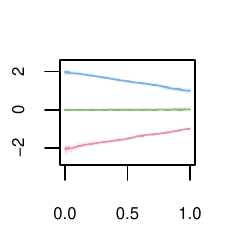} & \includegraphics[trim=0mm 0mm 0mm 5mm,clip]{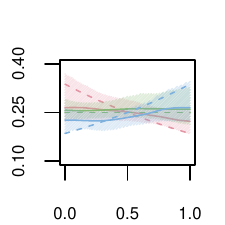} & \includegraphics[trim=0mm 0mm 0mm 5mm,clip]{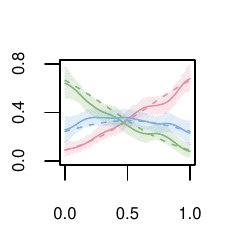} \end{tabular} }\\[5mm]
   \multicolumn{3}{c}{\textbf{Example 3}}\\[2mm] Mean & S.D. & Weights\\ \includegraphics[trim=0mm 0mm 0mm 0mm,clip]{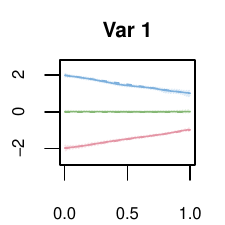} & \includegraphics[trim=0mm 0mm 0mm 0mm,clip]{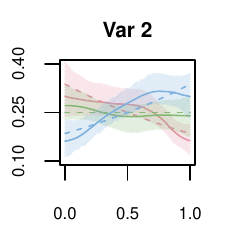} & \includegraphics[trim=0mm 0mm 0mm 0mm,clip]{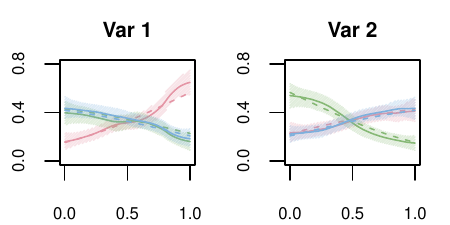} \end{tabular}
    \caption{Partial dependence plots for the three simulated examples based on the mixture model in (\ref{simex}). Example 1 (top row), Example 2 (middle row) and Example 3 (bottom row). Colours indicate the mixture component: red (1), green (2), and blue (3).The posterior median is shown by the solid line and the 95\% credible interval by the shaded region. The true function is shown as a dashed line.} \label{f:simulated_all_plots} 
    \end{figure}

Figure \ref{profs} shows the profiles of the functions we use for the component unnormalized weights, means and variances for Examples 2 and 3. This is to facilitate comparisons with the PDPs in Figure~\ref{f:simulated_all_plots}, which display the posterior medians of $\mu_k(\boldsymbol{x})$, $\sigma^2_k(\boldsymbol{x})$ and $w_k(\boldsymbol{x})$ of the SoftBART-CoRM's components means, variances and weights respectively.

\begin{table}[h!]
\begin{center}
\begin{tabular}{|l||c|c|c||c|c|c||c|c|c|}\hline 
& \multicolumn{3}{c||}{Example 1} & \multicolumn{3}{c||}{Example 2}& \multicolumn{3}{c|}{Example 3} \\
&\multicolumn{3}{c||}{Component} &\multicolumn{3}{c||}{Component} &\multicolumn{3}{c|}{Component}\\ 
& 1 & 2 & 3 & 1 & 2 & 3 & 1 & 2 & 3\\\hline 
Weight  & 0.40 & 0.39 & 0.18 & 0.35 & 0.32 & 0.32  & 0.34 & 0.33 & 0.33\\\hline 
$\kappa_k^{\delta}$ (mean)  & 0.03 & 0.02 & 0.02 & 0.1 & 0.01 & 0.08 & 0.09 & 0.01 & 0.09\\\hline 
$\kappa_k^{\phi}$ (standard deviation)  & 0.02 & 0.03 & 0.10 & 0.27 & 0.16 & 0.3 & 0.83 & 0.39 & 0.85\\\hline 
$\kappa_k^{\lambda}$ (weight)  & 0.06 & 0.06 & 0.03 & 1.07 & 1.35 & 0.46  & 1.24 & 1.06 & 1.34\\\hline 
\end{tabular}
\end{center}
  \caption{Simulated Examples: Averages weights and posterior means of $\kappa_k^{\delta}$, $\kappa_k^{\phi}$ and $\kappa_k^{\lambda}$  for the mean, standard deviation and weights respectively.}\label{t:simulated_variances}
\end{table}

The first step of our interpretation strategy is to assess if the SoftBART-CoRM can detect the presence (or absence) of predictor-driven variation in
each element of the mixture model in \eqref{simex}.
Table~\ref{t:simulated_variances} displays the posterior mean of the weights (averaged over iterations and observations), and the posterior medians of $\kappa^{\delta}_k$ (mean), $\kappa_k^{\phi}$ (variance) and $\kappa_k^{\lambda}$ (weight). The component weights and  the importance of each element of the mixture are correctly inferred for each example. 
In example 1, the $\kappa$'s are all near zero. In examples 2 and 3, the posterior medians of $\kappa_k^{\delta}$ and $\kappa_k^{\phi}$  are smaller for component 2, where the true mean and variance are constant, than for components 1 and 3, where both quantities vary with the predictor. In contrast, the posterior medians of $\kappa_k^{\lambda}$ are large across all components, correctly indicating predictor dependence in the mixture weights, in both examples.

\begin{table}[h!]
\begin{center}
\begin{tabular}{|l|c|c|c|}\hline 
Variable & Mean $\boldsymbol{s}^{\delta}$ & Variance $\boldsymbol{s}^{\phi}$& Weights $\boldsymbol{s}^{\lambda}$ \\
\hline Variable 1 & 0.99 & 0.04 & 0.55\\ Variable 2 & 0.01 & 0.96 & 0.45\\
\hline 
\end{tabular}
  \end{center}
  \caption{Simulated Example 3: Posterior means of $\boldsymbol{s}^{\delta}$ (mean), $\boldsymbol{s}^{\phi}$ (standard deviation) and $\boldsymbol{s}^{\lambda}$ (weights).}\label{t:simulated_VI}
\end{table}
We proceed to the second step, identification of important predictors, for Example 3 only since it has $p=2$ predictors and consider the posterior mean of the predictor importance parameters $\boldsymbol{s}^{\delta}$ (mean),  $\boldsymbol{s}^{\phi}$ (variance) and  $\boldsymbol{s}^{\lambda}$ (weights) shown in Table~\ref{t:simulated_VI}. These 
 clearly identify the importance of  $x_1$  for the mean, $x_2$ for the variance, and both variables for the mixture weights. These results demonstrate that SoftBART-CoRM can accurately determine not only whether a model component depends on predictors, but also which predictors are relevant for each element.

The partial dependency plots for all examples are  displayed in Figure~\ref{f:simulated_all_plots}. These show that the model correctly recovers the shapes of the weight, mean and variance functions across the predictor space. In example 1, the true finite mixture model where predictor effects are absent is recovered.
 The posterior medians of $\mu_k(\boldsymbol{x})$ and $w_k(\boldsymbol{x})$ (solid lines) are virtually indistinguishable from the corresponding true functions (dashed lines), with narrow 95\% credible intervals, for all components. Although the credible intervals for $\sigma^2_k(\boldsymbol{x})$ are slightly wider, its posterior median remains very close to the true variance function $\sigma^2(\boldsymbol{x})$. 
 In example 2, the posterior medians of all functions  closely track the corresponding true functions with narrow 95\% credible intervals, which  demonstrates that SoftBART-CoRM can  accurately estimate the functional form of predictor effects across the mixture weights, means, and variances, highlighting its ability to flexibly characterise complex distributional heterogeneity. In example 3, we shown PDPs only for the important variables, in all cases, the model is able to estimate  the underlying predictor effects.

\subsection{Wind Turbine Operational Data}\label{wind}

We apply the SoftBART-CORM model to the dataset \verb+windpw+ available in the R package \verb+kernplus+.  This is a popular data set for testing nonparametric kernel density regression models and assessing their ability to capture the non-linear effects of the predictors on  wind power output. The dataset contains 1000 observations,
the response is normalised power output relative to the rated power (\%) and there are five predictors: wind speed (m/s), wind direction (degrees), air density (kg/m$^3$), turbulence intensity, and below-hub wind shear (vertical wind profile).
\cite{LDGX015} discuss this data in more detail. 
We use the same MCMC set-up as the simulated examples.


\begin{table}[h!]
\small
\begin{center}
 \begin{tabular}{|l|c|c|c|c|c|}\hline 
  &\multicolumn{5}{c|}{Component}\\ & 1 & 2 & 3 & 4 & 5\\\hline 
  Weight  & 0.48 & 0.24 & 0.13 & 0.09 & 0.06\\\hline 
  $\kappa_k^{\delta}$ (mean)  & 0.7 & 0.61 & 1.57 & 0.12 & 0.38\\\hline
  $\kappa_k^{\phi}$ (standard deviation) & 1.54 & 0.98 & 0.2 & 5.11 & 0.3\\ \hline
  $\kappa_k^{\lambda}$ (weight)  & 10.87 & 16.02 & 30.99 & 18.99 & 4.59\\\hline
  \end{tabular}\\
  \end{center}
 \caption{Wind Turbine Data: Average posterior mean weights and posterior medians of $\kappa^\delta_k$, $\kappa^{\phi}_k $  and $\kappa^{\lambda}_k$ for the component means, standard deviations and weights respectively.}\label{t:wind_overall}
 \end{table}

Table~\ref{t:wind_overall} reports the non-negligible component weights and the posterior medians of $\kappa^\delta_k$, $\kappa^\phi_k$, and $\kappa^\lambda_k$.
Although five components are identified, we concentrate on the first four component which 
account for approximately $94\%$ of the total weight. The results indicate substantial predictor-driven variation in all four component means, with component~3 exhibiting the largest value of $\kappa^\delta_k$, followed by components~2 and~1. predictor effects on the component standard deviations are strongest for components~4, 1, and~2, as reflected by their large values of $\kappa^\phi_k$. In addition, the consistently large values of $\kappa^\lambda_k$ provide strong evidence of predictor dependence in the component weights across all four components.

  \begin{table}[h!]
 \small
 \begin{center}
  \vspace{0.1in} 
  \begin{tabular}{|l|c|c|c|}\hline 
Variable & Mean $\boldsymbol{s}^{\delta}$ & Variance $\boldsymbol{s}^{\phi}$& Weights $\boldsymbol{s}^{\lambda}$\\
  \hline 
  Wind Speed &  {0.76} &  {0.74}  & 0.32\\
  Wind Direction & 0.06 & 0.08 & 0.59\\
  Air Density & 0.05 & 0 & 0.03\\
  Turbulence Intensity & 0.07 & 0.17 & 0.04\\
  Below-Hub Wind Shear & 0.08 & 0 & 0.02\\\hline
  \end{tabular}
    \end{center}
\caption{Wind Turbine Data: Posterior means of $\boldsymbol{s}^{\delta}$ (mean), $\boldsymbol{s}^{\phi}$ (std deviation) and $\boldsymbol{s}^{\lambda}$ (weights).}\label{t:wind_VI}
\end{table}

Table~\ref{t:wind_VI} reports the posterior means of $\boldsymbol{s}^{\delta}$, $\boldsymbol{s}^{\phi}$, and $\boldsymbol{s}^{\lambda}$, the predictor importance probabilities, where larger values indicate stronger predictor effects. Wind speed is the most influential predictor overall, particularly for the component means and standard deviations. Wind direction predominantly governs variation in the mixture weights, whereas turbulence intensity mainly affects the component standard deviations.

\begin{figure}[h!]
\begin{center}
\begin{tabular}{c}
Mean\\
\includegraphics{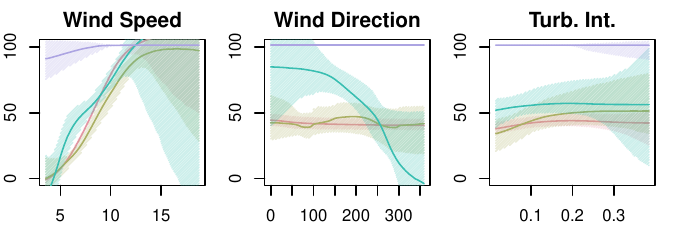}\\
Standard Deviation\\
\includegraphics{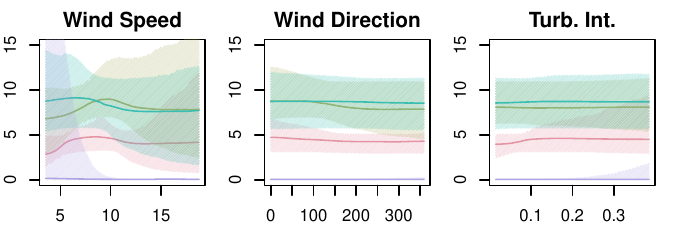}\\
Weights\\
\includegraphics{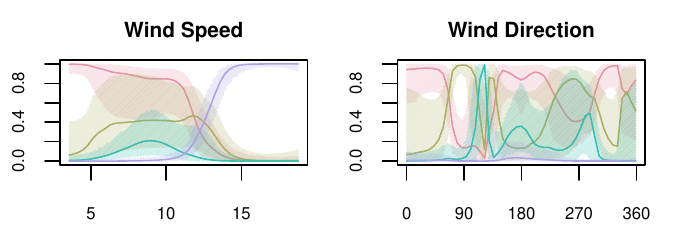}\\
\end{tabular}
\end{center}
\caption{Wind Turbine Data: Posterior median partial dependence plots (solid lines) with 95\% credible intervals (shaded regions) for the component means (top), standard deviations (middle) and weights (bottom). Lines and regions are coloured according the component: red (1), brown (2), turquoise (3) and purple (4). }\label{f:wind_effects}
\end{figure}

Figure~\ref{f:wind_effects} displays the posterior median partial dependence plots (solid lines) for the most influential predictors in the component-specific means, standard deviations, and mixing weights, together with 95\% credible intervals (shaded regions). The most pronounced predictor effects are observed in the mixing weights and component means. Component~4 (purple) corresponds to a high wind-speed regime. Its mean function is relatively insensitive to wind speed beyond approximately 12\,m/s, indicating that normalised power has reached a saturation level in this region. In contrast, allocation to the remaining mixture components is driven primarily by wind direction. Component~1 (red) has the largest mixing weight for wind directions between $0^\circ$ and $80^\circ$,$150^\circ$ and $210^\circ$, and between $290^\circ$ and $350^\circ$. Component~2 (brown) dominates over the ranges $70^\circ$--$150^\circ$ and $250^\circ$--$290^\circ$, whereas Component~3 (turquoise) is associated with a localised directional regime, receiving substantial weight between $105^\circ$ and $135^\circ$. While Components~1 and~2 have similar mean functions, the component-specific standard deviation of Component~2 is approximately twice that of Component~1. The standard deviation of Component~3 is similar to that of Component~2, but it has a consistently higher mean across the range of wind speeds. Overall, the mixing weights reveal that wind direction guides component allocation, while the component-specific mean and variance functions capture distinct operating regimes with differing levels of power output and variability.


 

\subsection{Temperature prediction}\label{temp}

The second application looks at   predicting the temperature (measured in celsius) at noon at a location near Monash University, Clayton, Victoria, Australia from 1st January 2010 to 31st May 2021. The dataset includes 4169 observations and is available from 
\url{https://www.kaggle.com/datasets/nirmalsankalana/oikolab-weather-dataset}. We study the effect of eight predictors on midday temperature, which are listed in Table~\ref{t:weather_VI}. The MCMC set up from the simulated examples is used.

\begin{table}[h!]
{\small
\begin{center}
 \begin{tabular}{|l|c|c|c|}\hline 
 &\multicolumn{3}{c|}{Component}\\ & 1 & 2 & 3\\\hline 
 Weight  & 0.35 & 0.34 & 0.31\\\hline
 $\kappa_k^{\delta}$ (mean)  & 0.41 & 0.85 & 0.42\\ \hline
 $\kappa_k^{\phi}$ (standard deviation) & 0.18 & 0.82 & 1.28\\ \hline
 $\kappa_k^{\lambda}$ (weight)  & 9.37 & 12.47 & 14.07\\\hline 
 \end{tabular}\\
 \end{center}}
 \caption{Temperature Prediction: Average posterior mean weights and posterior medians of $\kappa^\delta_k$, $\kappa^{\phi}_k $  and $\kappa^{\lambda}_k$ for the mean, standard deviation and weights respectively. }\label{t:weather_overall}
 \end{table}


Table~\ref{t:weather_overall} summarizes the posterior mean component weights together with the posterior medians of $\kappa^\delta_k$, $\kappa^\phi_k$, and $\kappa^\lambda_k$. 
The three components identified have weights of similar magnitude $\approx \frac{1}{3}$. The high estimated scale parameters across all three components indicate strong predictor effects on both the component means and the mixing weights. predictor effects on the component standard deviations are also substantial for Components~2 and~3, whereas the standard deviation of Component~1 exhibits weak predictor dependence.

\begin{table}[h!]
 {\small
 \begin{center}
 \vspace{0.2in} 
 \begin{tabular}{|l|l|c|c|c|}\hline 
 Name & Variable & Mean $\boldsymbol{s}^{\delta}$ & Variance $\boldsymbol{s}^{\phi}$& Weights $\boldsymbol{s}^{\lambda}$ \\ \hline
Wind Speed &  Wind Speed &  { 0.15} &  { 0.43}  & 0.11\\
 SLP & Mean Sea Level Pressure  & 0 & 0 & 0.16\\
 Rel. Humidity & Relative Humidity & 0.08 & 0 & 0.2\\
 STR &  Surface Thermal Radiation & 0.37 & 0.33 & 0.24\\
 Cloud Cover & Cloud Cover & 0.09 &  0.23 & 0.11\\
 Month & Month & 0 & 0 & 0\\
 Week & Week & 0.07 & 0 & 0\\
 Quarter & Quarter & 0.25 & 0 & 0.18\\\hline
 \end{tabular}
 \end{center}}
 \caption{Temperature Prediction: Posterior means of $\boldsymbol{s}^{\delta}$ (mean), $\boldsymbol{s}^{\phi}$ (standard deviation) and $\boldsymbol{s}^{\lambda}$ (weights).}\label{t:weather_VI}
\end{table}

Table~\ref{t:weather_VI} reports the posterior means of $\boldsymbol{s}^{\delta}$, $\boldsymbol{s}^{\phi}$, and $\boldsymbol{s}^{\lambda}$. With the exception of Month and Week, all predictors influence at least one element of the SoftBART-CoRM model. Wind Speed and Surface Thermal Radiation (STR), and to a lesser extent Cloud Cover, are important drivers of the component means, standard deviations, and mixing weights. Quarter additionally contributes to variation in the component means,  and mixing weights, while Mean Sea Level Pressure (SLP), Relative Humidity, and Quarter are influential predictors of the mixing weights. 

\begin{figure}[h!]
\begin{center}
Mean\\
\includegraphics{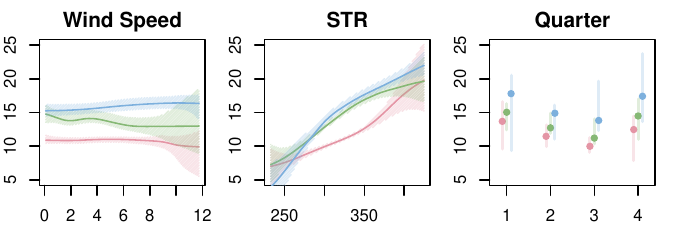}\\
Standard Deviation\\
\includegraphics{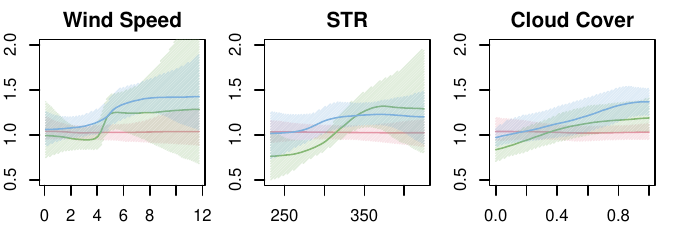}
\end{center}
\caption{Temperature Prediction: Posterior median partial dependence plots (solid line) and 95\% credible intervals (shaded regions) for the most influential predictors in the component-specific mean and standard deviation functions. The colours denote Components~1 (red), 2 (green), and 3 (blue).}\label{f:weather_effects_1}
\end{figure}

\begin{figure}[h!]
\begin{center}
\begin{tabular}{c}
\includegraphics{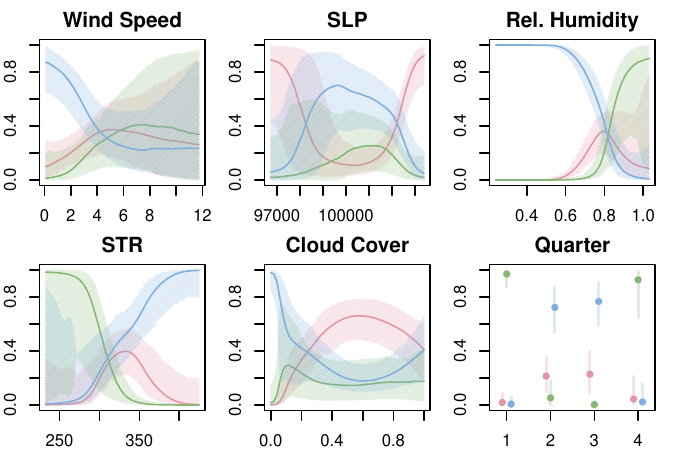}
\end{tabular}
\end{center}
\caption{Temperature Prediction: Posterior median partial dependence plots (solid lines) and 95\% credible intervals (shaded regions) of the most influential variables for the weight in the three components. The colours denote Components~1 red , 2 (green) and 3 (blue). }\label{f:weather_effects_2}
\end{figure}

Figures~\ref{f:weather_effects_1} and~\ref{f:weather_effects_2} show the posterior partial dependence plots (solid lines) with 95\% credible intervals (shaded regions) of the most influential predictors for the component means, standard deviations and weights. The component-specific mean functions reveal three distinct midday temperature regimes, with the overall temperature level increasing from Component~1 to Component~3 for the three predictors. STR has a strong positive effect on the component means, although the form of this relationship differs across components. In particular, Component~1 exhibits lower median temperatures for intermediate values of STR, whereas the response profiles of Components~2 and~3 are similar. The standard deviation functions further differentiate the components. Consistent with the small value of $\kappa^\phi_1$, the standard deviation of Component~1 is insensitive to the predictors. In contrast, the standard deviations of Components~2 and~3 increase with Wind Speed, STR, and Cloud Cover, indicating greater heteroscedasticity within these regimes. The mixing weights exhibit strong predictor dependence. Component~2 dominates during Quarters~1 and~4, with mixing weights close to one, while Components~1 and~3 receive relatively little mass. Component~3 is most heavily favoured under conditions of low Wind Speed, Relative Humidity, STR and Cloud Cover, together with moderate-to-high values of SLP. Overall, the results indicate that meteorological conditions influence temperature through both regime membership and regime-specific response characteristics.


\subsection{Predictive Performance}

We compare the out-of-sample predictive performance of the SoftBART model to three benchmark models  (linear regression, SoftBART, random forest  \citep{Bre::2001} and XGBoost  \citep{Che:2016}) 
and three extensions of the SoftBART model:
\begin{itemize}
\item mix -- A mixture of SoftBART regression models -- $p(y \mid \boldsymbol{x}) = \sum_{k=1}^K w_k\, \mbox{N}(y \mid \mu_k(\boldsymbol{x}), \sigma^2_k)$
where $\mu_1(\boldsymbol{x}), \dots, \mu_K(\boldsymbol{x})$ are given independent SoftBART priors, which allows for different regressions in the mean of each component only.
\item split -- A SoftBART model with Dirichlet process mixture errors --  $y_i  = \mu(\boldsymbol{x}_i) + \epsilon_i$ where $\mu(\boldsymbol{x}_i)$ is given a SoftBART prior and $\epsilon_i\sim F$ where $F$ is given a Dirichlet process mixture of normal distributions (which uses  the default prior of \cite{gri10}). 
\item  hetero -- A heteroscedastic SoftBART regression model -- $ p(y \mid \boldsymbol{x}) = \mbox{N}(y \mid \mu(\boldsymbol{x}), \sigma^2(\boldsymbol{x}))$ where 
$\mu(\boldsymbol{x})$ and
$\log \sigma^2(\boldsymbol{x}_i)$ are given independent SoftBART priors. 
\end{itemize}
Performance is evaluated using the root mean squared error (RMSE) and the log predictive score (LPS) with 10-fold cross validation on both the wind turbine and temperature.

\begin{table}[h!]
\begin{center}
\begin{tabular}{|c|c|ccc|ccccc|}\hline
& & Linear & Random & XG & \multicolumn{5}{c|}{SoftBART} \\
& & Reg. & Forest & Boost & orginal & mix & split & hetero& CORM  \\\hline
Wind & RMSE & 14.75 & 11.22 & 10.62 & 9.62 & 11.20 & 9.50 &12.77 & {\bf 9.44}\\
Turbine & LPS & -4.11 & -3.84 & -10.07 & -3.67 & -4.63 & -3.66 & -3.43 & {\bf -3.15}\\\hline
Temperature & RMSE & 2.02 & 1.63 & 1.56 & 1.51 & 1.59 & 1.54 & 1.56 & {\bf 1.46}\\
& LPS & -2.12 & -1.91 & -2.10 & -1.83 & -3.30 & -1.85 & -1.79 & {\bf -1.69}\\\hline
\end{tabular}
\end{center}
\caption{Root mean squared error (RMSE) and log predictive score (LPS) from 10-fold cross-validation  for the Wind Turbine Operational and Temperature Prediction Data.}\label{t:CV_results}
\end{table}


The cross-validation results, displayed in Table~\ref{t:CV_results}, are fairly consistent across the two data sets. All three variants of SoftBART (original, split, and CoRM) achieve lower RMSE values than the competing regression methods (Linear Regression, Random Forest, and XGBoost), with SoftBART-CoRM providing a modest improvement over the original SoftBART. In contrast, the mixture-of-regressions (mix) and heteroscedastic (hetero) SoftBART formulations exhibit noticeably weaker performance, comparable to that of the standard regression approaches. The most substantial differences, however, are observed in the log predictive scores (LPS). SoftBART-CoRM achieves the best performance across both data sets, improving on the original SoftBART by 0.52 for the wind turbine data and 0.14 for the temperature data. The split SoftBART with DPM errors performs similarly to the original SoftBART, while the heteroscedastic SoftBART yields some improvement in LPS but remains considerably inferior to SoftBART-CoRM. With the exception of the mixture-of-regressions variant, all SoftBART-based models outperform the standard regression approaches in terms of predictive density estimation. These results illustrate the importance of allowing predictor dependence in both the weights and the kernel parameters particularly for accurately modelling the whole conditional distribution.

\section{Discussion}

The SoftBART-CoRM is a highly flexible Bayesian density regression model which uses the SoftBART prior to allow non-linear effects in the mean, variance and weights of an infinite mixture model and avoids overfitting by shrinking towards simpler models. The use of the DART prior also mitigates against overfitting and provides a simple method for variable selection. The simulated examples show that the model is able to both correctly identify important predictors and correctly estimate the relationship between the predictors and the response. This leads to an approach to interpreting the model that emphasizes the relationship between predictors and the different elements of the model. 
The particularly strong LPS performance of SoftBART-CoRM highlights the benefits of its modelling strategy. Unlike the original SoftBART, which relies on a single regression function and a relatively simple residual structure, SoftBART-CoRM models the conditional response distribution through a predictor-dependent mixture whose component means, variances, and mixing weights are all allowed to vary flexibly with the predictors. This construction enables the model to adapt simultaneously to nonlinear mean effects, heterogeneous variances, skewness, multi-modality, and changes in component prevalence across the predictor space. Consequently, SoftBART-CoRM is able to produce predictive distributions that are substantially better calibrated and more closely aligned with the underlying data-generating mechanisms. The marked improvements in LPS therefore suggest that the primary advantage of SoftBART-CoRM lies not merely in improved predictive accuracy, but in its ability to accurately characterise complex conditional distributions and deliver substantially more realistic out-of-sample predictive densities.

The paper has focussed on problems where the data has a few predictors (1 to 10) and a moderate number of observations (up to a few thousand). Future work will consider scaling the model in both the number of predictors and the number of observations. The SoftBART prior, which includes a predictor importance prior, is designed to have good performance  with a substantially larger number of variables and this will allow the SoftBART-CORM to be applied to problems with many more regressors without substantial changes to the model and the MCMC implementation. As with many MCMC methods for mixture models, scaling to a substantially larger $n$ is more challenging. One approach would be to build on the well-established variational inference methods for mixture models
\citep{BlKuMc17}. There is a much smaller literature on developing variational inference approaches for BART-type models but there has been some recent work 
\citep[see e.g.][]{Salazar23}. Another direction of future work would be to use more general kernels, such as skew-normal or skew-$t$ distributions. There has been some work on using more general distributions than the normal with BART-type models  \citep[e.g.][]{PrChGeMc20, UmLiSiBa23} but there still substantial work to do.
For example, does allow the shape parameters to depend on predictors. As we have shown, allowing the different elements of a mixture model to depend on predictors can be key to good predictive performance. Using more flexible kernel distributions can lead to more parsimonious models which use less components and can provide more easily understood inference about the effect of predictors on features of the distribution such as skewness or heavy tails.



 \bibliographystyle{chicago}
\bibliography{ref}

\begin{thebibliography}{}

\bibitem[\protect\citeauthoryear{Barrientos, Jara, and Quintana}{Barrientos
  et~al.}{2012}]{BaJaQu12}
Barrientos, A.~F., A.~Jara, and F.~A. Quintana (2012).
\newblock On the support of {M}ac{E}achern's {D}ependent {D}irichlet
  {P}rocesses and extensions.
\newblock {\em Bayesian Analysis\/}~{\em 7}, 277--310.

\bibitem[\protect\citeauthoryear{Bernaciak and Griffin}{Bernaciak and
  Griffin}{2026}]{BerGri26}
Bernaciak, D. and J.~E. Griffin (2026).
\newblock {A General-Purpose Approximation to the Ferguson-Klass Algorithm for
  Sampling from L{\'e}vy Processes Without Gaussian Components}.
\newblock {\em Journal of Computational and Graphical Statistics\/}~{\em 35},
  1140--1150.

\bibitem[\protect\citeauthoryear{Blei, Kucukelbir, and McAuliffe}{Blei
  et~al.}{2017}]{BlKuMc17}
Blei, D.~M., A.~Kucukelbir, and J.~D. McAuliffe (2017).
\newblock {Variational Inference: A Review for Statisticians}.
\newblock {\em Journal of the American Statistical Association\/}~{\em 112},
  859--877.

\bibitem[\protect\citeauthoryear{Breiman}{Breiman}{2001}]{Bre::2001}
Breiman, L. (2001).
\newblock Random forests.
\newblock {\em Machine Learning\/}~{\em 45}, 5---32.

\bibitem[\protect\citeauthoryear{Chen and Guestrin}{Chen and
  Guestrin}{2016}]{Che:2016}
Chen, T. and C.~Guestrin (2016).
\newblock {XGBoost}: A scalable tree boosting system.
\newblock In {\em Proceedings of the 22nd ACM SIGKDD International Conference
  on Knowledge Discovery and Data Mining}, pp.\  785---794.

\bibitem[\protect\citeauthoryear{Chernozhukov, Fernandez-Val, and
  Blaise}{Chernozhukov et~al.}{2013}]{chern2013}
Chernozhukov, V., I.~Fernandez-Val, and M.~Blaise (2013).
\newblock Infernece on {C}ounterfactual {D}istributions.
\newblock {\em Econometrica\/}~{\em 81\/}(6), 2205--2268.

\bibitem[\protect\citeauthoryear{Chipman, George, and McCulloch}{Chipman
  et~al.}{2010}]{ChiGeoMcc10}
Chipman, H.~A., E.~I. George, and R.~E. McCulloch (2010).
\newblock {BART: Bayesian additive regression trees}.
\newblock {\em Annals of Applied Statistics\/}~{\em 4}, 266--298.

\bibitem[\protect\citeauthoryear{De~Iorio, M{\"u}ller, Rosner, and
  MacEachern}{De~Iorio et~al.}{2004}]{iorio04}
De~Iorio, M., P.~M{\"u}ller, G.~L. Rosner, and S.~N. MacEachern (2004).
\newblock An {A}{N}{O}{V}{A} {M}odel for {D}ependent {R}andom {M}easures.
\newblock {\em Journal of the American Statistical Assocation\/}~{\em 99},
  205--215.

\bibitem[\protect\citeauthoryear{Dorie, Chipman, and McCulloch}{Dorie
  et~al.}{2026}]{DoChMc26}
Dorie, V., H.~Chipman, and R.~McCulloch (2026).
\newblock {\em dbarts: Discrete Bayesian Additive Regression Trees Sampler}.
\newblock R package version 0.9-34.

\bibitem[\protect\citeauthoryear{Dunson and Park}{Dunson and
  Park}{2008}]{dunson08}
Dunson, D.~B. and J.-H. Park (2008).
\newblock Kernel stick-breaking processes.
\newblock {\em Biometrika\/}~{\em 95}, 307--323.

\bibitem[\protect\citeauthoryear{Friedman}{Friedman}{2001}]{Fri01}
Friedman, J.~H. (2001).
\newblock Greedy function approximation: A gradient boosting machine.
\newblock {\em Annals of Statistics\/}~{\em 29}, 1189--1232.

\bibitem[\protect\citeauthoryear{George, Laud, Logan, McCulloch, and
  Sparapani}{George et~al.}{2019}]{GeLaLoMcSp19}
George, E., P.~Laud, B.~Logan, R.~McCulloch, and R.~Sparapani (2019).
\newblock {Fully Nonparametric Bayesian Additive Regression Trees}.
\newblock In {\em Topics in Identification, Limited Dependent Variables,
  Partial Observability, Experimentation, and Flexible Modeling: Part B}.
  Emerald Publishing Limited.

\bibitem[\protect\citeauthoryear{Ghosal, Ghosh, and Ramamoorthi}{Ghosal
  et~al.}{1999}]{Ghosal1999}
Ghosal, S., J.~K. Ghosh, and R.~V. Ramamoorthi (1999).
\newblock Posterior {C}onsistency of {D}irichlet {M}ixtures in {D}ensity
  {E}stimation.
\newblock {\em The Annals of Statistics\/}~{\em 27\/}(1), 143--158.

\bibitem[\protect\citeauthoryear{Ghosal, Ghosh, and van~der Vaart}{Ghosal
  et~al.}{2001}]{Ghosal2001}
Ghosal, S., J.~K. Ghosh, and A.~van~der Vaart (2001).
\newblock Entropy and {P}rior {M}ass {C}onditions for {P}osterior
  {C}onsistency.
\newblock {\em The Annals of Statistics\/}~{\em 28\/}(2), 500--531.

\bibitem[\protect\citeauthoryear{Ghosal and van~der Vaart}{Ghosal and van~der
  Vaart}{2007}]{Ghosal2007}
Ghosal, S. and A.~van~der Vaart (2007).
\newblock Convergence {R}ates of {P}osterior {D}istributions for {N}on iid
  {O}bservations.
\newblock {\em The Annals of Statistics\/}~{\em 35\/}(1), 192--223.

\bibitem[\protect\citeauthoryear{Griffin}{Griffin}{2010}]{gri10}
Griffin, J.~E. (2010).
\newblock Default priors for density estimation with mixture models.
\newblock {\em Bayesian Analysis\/}~{\em 5}, 45--64.

\bibitem[\protect\citeauthoryear{Griffin and Leisen}{Griffin and
  Leisen}{2017}]{GriLei17}
Griffin, J.~E. and F.~Leisen (2017).
\newblock {Compound Random Measures and their Use in Bayesian Non-Parametrics}.
\newblock {\em Journal of the Royal Statistical Society Series B\/}~{\em 79},
  525--545.

\bibitem[\protect\citeauthoryear{Griffin and Leisen}{Griffin and
  Leisen}{2018}]{GriLei18}
Griffin, J.~E. and F.~Leisen (2018).
\newblock Modelling and computation using {NCoRM} mixtures for density
  regression.
\newblock {\em Bayesian Analysis\/}~{\em 13}, 897--916.

\bibitem[\protect\citeauthoryear{Griffin and Steel}{Griffin and
  Steel}{2006}]{griste06}
Griffin, J.~E. and M.~F.~J. Steel (2006).
\newblock {Order-Based Dependent Dirichlet Processes}.
\newblock {\em Journal of the American Statistical Association\/}~{\em 101},
  179--193.

\bibitem[\protect\citeauthoryear{Hill, Linero, and Murray}{Hill
  et~al.}{2020}]{hill20}
Hill, J., A.~Linero, and J.~Murray (2020).
\newblock Bayesian {A}dditive {R}egression {T}rees: {A} {R}eview and {L}ook
  {F}orward.
\newblock {\em Annual Review of Statistics and Its Applications\/}~{\em 7},
  251--278.

\bibitem[\protect\citeauthoryear{Hothorn, Kneib, and B\"uhlmann}{Hothorn
  et~al.}{2014}]{Hothorn2014}
Hothorn, T., T.~Kneib, and P.~B\"uhlmann (2014).
\newblock Conditional {T}ransformation {M}odels.
\newblock {\em Journal of the Royal Statistical Society, Series B\/}~{\em
  76\/}(76), 3--27.

\bibitem[\protect\citeauthoryear{Kim, Shephard, and Chib}{Kim
  et~al.}{1998}]{KSC98}
Kim, S., N.~Shephard, and S.~Chib (1998).
\newblock {Stochastic Volatility: Likelihood Inference and Comparison with ARCH
  Models}.
\newblock {\em The Review of Economic Studies\/}~{\em 65}, 361--393.

\bibitem[\protect\citeauthoryear{Kindo, Wang, Hanson, and Pe{\~n}a}{Kindo
  et~al.}{2016}]{KiWaHaPe16}
Kindo, B.~P., H.~Wang, T.~Hanson, and E.~A. Pe{\~n}a (2016).
\newblock Bayesian quantile additive regression trees.
\newblock https://arxiv.org/abs/1607.02676.

\bibitem[\protect\citeauthoryear{Klein}{Klein}{2024}]{Klein2024}
Klein, N. (2024).
\newblock Distributional {R}egression for {D}ata {A}nalysis.
\newblock {\em Annual Review of Statistics and Its Applications\/}~{\em 11},
  321--346.

\bibitem[\protect\citeauthoryear{Klein, Kneib, Lang, and Sohn}{Klein
  et~al.}{2015}]{Klein2015}
Klein, N., T.~Kneib, S.~Lang, and A.~Sohn (2015).
\newblock Bayesian {S}tructured {A}dditive {D}istributional {R}egression.
\newblock {\em Journal of the Royal Statistical Society, Series C\/}~{\em
  64\/}(4), 569--591.

\bibitem[\protect\citeauthoryear{Koenker and Basset}{Koenker and
  Basset}{1978}]{Koenker1978}
Koenker, R. and G.~Basset (1978).
\newblock Regression {Q}uantiles.
\newblock {\em Econometrica\/}~{\em 46\/}(1), 33--50.

\bibitem[\protect\citeauthoryear{Lee, Ding, Genton, and Xie}{Lee
  et~al.}{2015}]{LDGX015}
Lee, G., Y.~Ding, M.~G. Genton, and L.~Xie (2015).
\newblock Power curve estimation with multivariate environmental factors for
  inland and offshore wind farms.
\newblock {\em Journal of the American Statistical Association\/}~{\em 110},
  56--67.

\bibitem[\protect\citeauthoryear{Li, Linero, and Murray}{Li
  et~al.}{2023}]{LiLiMu23}
Li, Y., A.~R. Linero, and J.~Murray (2023).
\newblock {Adaptive Conditional Distribution Estimation with Bayesian Decision
  Tree Ensembles}.
\newblock {\em Journal of the American Statistical Assocation\/}~{\em 118},
  2129--2143.

\bibitem[\protect\citeauthoryear{Lim, Lu, St.~Ville, and Chen}{Lim
  et~al.}{2025}]{LiLuSVCh25}
Lim, Y., R.~Lu, M.~St.~Ville, and Z.~Chen (2025).
\newblock {Bayesian additive tree ensembles for composite quantile
  regressions}.
\newblock {\em Statistics and Computing\/}~{\em 35}, 175.

\bibitem[\protect\citeauthoryear{Linero}{Linero}{2018}]{Lin18}
Linero, A.~R. (2018).
\newblock {Bayesian regression trees for high-dimensional prediction and
  variable selection}.
\newblock {\em Journal of the American Statistical Association\/}~{\em 113},
  626--636.

\bibitem[\protect\citeauthoryear{Linero}{Linero}{2022}]{Li22}
Linero, A.~R. (2022).
\newblock Softbart: Soft bayesian additive regression trees.

\bibitem[\protect\citeauthoryear{Linero and Yang}{Linero and
  Yang}{2018}]{LiYa18}
Linero, A.~R. and Y.~Yang (2018).
\newblock Bayesian regression tree ensembles that adapt to smoothness and
  sparsity.
\newblock {\em Journal of the Royal Statistical Society, Series B\/}~{\em 80},
  1087--1110.

\bibitem[\protect\citeauthoryear{Lo}{Lo}{1984}]{Lo1984}
Lo, A.~Y. (1984).
\newblock On the {C}lass of {B}ayesian {N}onparameteric {E}stimates: {I}.
  {D}ensity {E}stimates.
\newblock {\em The Annals of Statistics\/}~{\em 12\/}(1), 351--357.

\bibitem[\protect\citeauthoryear{MacEachern}{MacEachern}{1999}]{MacEachern1999}
MacEachern, S.~N. (1999).
\newblock Dependent {N}on {P}arametric {P}rocesses.
\newblock In {\em ASA Proceedings of the Section on Bayesian Statistical
  Science}, pp.\  50--55.

\bibitem[\protect\citeauthoryear{MacEachern}{MacEachern}{2000}]{Mac00}
MacEachern, S.~N. (2000).
\newblock Dependent {D}irichlet processes.
\newblock Technical report, Department of Statistics, Ohio State University.

\bibitem[\protect\citeauthoryear{Norets and Pati}{Norets and
  Pati}{2014}]{Norets2014}
Norets, A. and D.~Pati (2014).
\newblock {Adaptive Bayesian Estimation of Conditional Densities}.
\newblock {\em Econometric Theory\/}~{\em 30}, 939--977.

\bibitem[\protect\citeauthoryear{Omori, Chib, Shephard, and Nakajima}{Omori
  et~al.}{2007}]{OCSN07}
Omori, Y., S.~Chib, N.~Shephard, and J.~Nakajima (2007).
\newblock Stochastic volatility with leverage: {F}ast and efficient likelihood
  inference.
\newblock {\em Journal of Econometrics\/}~{\em 140}, 425--449.

\bibitem[\protect\citeauthoryear{Orlandi, Murray, Linero, and
  Volfovsky}{Orlandi et~al.}{2021}]{OrMuLiVo21}
Orlandi, V., J.~Murray, A.~Linero, and A.~Volfovsky (2021).
\newblock {Density Regression with Bayesian Additive Regression Trees}.
\newblock https://arxiv.org/abs/2112.12259.

\bibitem[\protect\citeauthoryear{Papastamoulis}{Papastamoulis}{2016}]{Pa16}
Papastamoulis, P. (2016).
\newblock {{label.switching}: An {R} Package for Dealing with the Label
  Switching Problem in MCMC Outputs}.
\newblock {\em Journal of Statistical Software, Code Snippets\/}~{\em 69\/}(1),
  1--24.

\bibitem[\protect\citeauthoryear{Pati, Dunson, and Tokdar}{Pati
  et~al.}{2013}]{Pati2013}
Pati, D., D.~B. Dunson, and S.~T. Tokdar (2013).
\newblock Posterior consistency in conditional distribution estimation.
\newblock {\em Journal of Multivariate Analysis\/}~{\em 116}, 456--472.

\bibitem[\protect\citeauthoryear{Polson, Scott, and Windle}{Polson
  et~al.}{2013}]{PSW13}
Polson, N.~G., J.~G. Scott, and J.~Windle (2013).
\newblock {Bayesian Inference for Logistic Models Using P\'olya-Gamma Latent
  Variables}.
\newblock {\em Journal of the American Statistical Association\/}~{\em 108},
  1339--1349.

\bibitem[\protect\citeauthoryear{Pratola, Chipman, George, and
  McCulloch}{Pratola et~al.}{2020}]{PrChGeMc20}
Pratola, M.~T., H.~A. Chipman, E.~I. George, and R.~E. McCulloch (2020).
\newblock {Heteroscedastic BART via Multiplicative Regression Trees}.
\newblock {\em Journal of Computational and Graphical Statistics\/}~{\em 29},
  405--417.

\bibitem[\protect\citeauthoryear{Quintana, M{\"u}ller, Jara, and
  MacEachern}{Quintana et~al.}{2022}]{quin22}
Quintana, F.~A., P.~M{\"u}ller, A.~Jara, and S.~N. MacEachern (2022).
\newblock The {D}ependent {D}irichlet {P}rocess and {R}elated {M}odels.
\newblock {\em Statistical Science\/}~{\em 37}, 24--41.

\bibitem[\protect\citeauthoryear{Rigby and Stasinopoulos}{Rigby and
  Stasinopoulos}{2005}]{Rigby2005}
Rigby, R.~A. and D.~M. Stasinopoulos (2005).
\newblock Generalised {A}dditive {M}odels for {L}ocation, {S}cale and {S}hape.
\newblock {\em Journal of the Royal Statistical Society, Series C\/}~{\em
  54\/}(3), 507--554.

\bibitem[\protect\citeauthoryear{Riva~Palacio and Leisen}{Riva~Palacio and
  Leisen}{2018}]{RPLe18}
Riva~Palacio, A. and F.~Leisen (2018).
\newblock Integrability conditions for compound random measures.
\newblock {\em Statistics and Probability Letters\/}~{\em 135}, 32--37.

\bibitem[\protect\citeauthoryear{Rodr{\'\i}guez and Dunson}{Rodr{\'\i}guez and
  Dunson}{2011}]{dunson11}
Rodr{\'\i}guez, A. and D.~Dunson (2011).
\newblock Nonparametric {B}ayesian models through probit stick-breaking
  processes stick-breaking processes.
\newblock {\em Bayesian Analysis\/}~{\em 6}, 145--178.

\bibitem[\protect\citeauthoryear{Rodriguez and Walker}{Rodriguez and
  Walker}{2014}]{RodWal14}
Rodriguez, C.~E. and S.~Walker (2014).
\newblock {Label Switching in Bayesian Mixture Models: Deterministic relabeling
  strategies}.
\newblock {\em Journal of Computational and Graphical Statistics\/}~{\em 23},
  25--45.

\bibitem[\protect\citeauthoryear{Rousseau and Mengersen}{Rousseau and
  Mengersen}{2011}]{RoMe11}
Rousseau, J. and K.~Mengersen (2011, 11).
\newblock Asymptotic behaviour of the posterior distribution in overfitted
  mixture models.
\newblock {\em Journal of the Royal Statistical Society Series B: Statistical
  Methodology\/}~{\em 73\/}(5), 689--710.

\bibitem[\protect\citeauthoryear{Ro\v{c}kov\'a and van~der Pas}{Ro\v{c}kov\'a
  and van~der Pas}{2020}]{RoVDP20}
Ro\v{c}kov\'a, V. and S.~van~der Pas (2020).
\newblock Posterior concentration for {B}ayesian regression trees and forests.
\newblock {\em Annals of Statistics\/}~{\em 48}, 2018--2131.

\bibitem[\protect\citeauthoryear{Sakaria and Griffin}{Sakaria and
  Griffin}{2017}]{SakGri17}
Sakaria, D.~K. and J.~E. Griffin (2017).
\newblock On efficient {B}ayesian inference for models with stochastic
  volatility.
\newblock {\em Econometrics and Statistics\/}~{\em 3}, 23--33.

\bibitem[\protect\citeauthoryear{Salazar}{Salazar}{2023}]{Salazar23}
Salazar, S. (2023).
\newblock Vart: Variational regression trees.
\newblock In A.~Oh, T.~Naumann, A.~Globerson, K.~Saenko, M.~Hardt, and
  S.~Levine (Eds.), {\em Advances in Neural Information Processing Systems},
  Volume~36, pp.\  45681--45693. Curran Associates, Inc.

\bibitem[\protect\citeauthoryear{Sparapani, Logan, Maiers, Laud, and
  McCulloch}{Sparapani et~al.}{2023}]{SpLoMaLaMc23}
Sparapani, R.~A., B.~R. Logan, M.~Maiers, P.~W. Laud, and R.~McCulloch (2023).
\newblock {Nonparametric failure time: Time-to-event machine learning with
  heteroskedastic Bayesian additive regression trees and low information
  omnibus Dirichlet process mixtures}.
\newblock {\em Biometrics\/}~{\em 79}, 3023--3037.

\bibitem[\protect\citeauthoryear{Tokdar and Ghosh}{Tokdar and
  Ghosh}{2007}]{Tokdar2007}
Tokdar, S.~T. and J.~K. Ghosh (2007).
\newblock {Posterior Consistency of Logistic Gaussian Process Priors in Density
  Estimation}.
\newblock {\em Journal of Planning and Inference\/}~{\em 137}, 34--42.

\bibitem[\protect\citeauthoryear{Um, Linero, Sinha, and Bandyopadhyay}{Um
  et~al.}{2023}]{UmLiSiBa23}
Um, S., A.~R. Linero, D.~Sinha, and D.~Bandyopadhyay (2023).
\newblock Bayesian additive regression trees for multivariate skewed responses.
\newblock {\em Statistics in Medicine\/}~{\em 42}, 246--263.

\bibitem[\protect\citeauthoryear{Umlauf, Klein, and Zeileis}{Umlauf
  et~al.}{2018}]{Umlauf2018}
Umlauf, N., N.~Klein, and A.~Zeileis (2018).
\newblock {B}{A}{M}{L}{S}{S}: {B}ayesian {A}dditive {M}odels for {L}ocation,
  {S}cale and {S}hape.
\newblock {\em Journal of Computational and Graphical Statisitics\/}~{\em
  27\/}(3), 612--627.

\bibitem[\protect\citeauthoryear{Wade and In\'acio}{Wade and
  In\'acio}{2025}]{WaIn25}
Wade, S. and V.~In\'acio (2025).
\newblock Bayesian {D}ependent {M}ixture {M}odels: {A} {P}redictive
  {C}omparison and {S}urvey.
\newblock {\em Statistical Science\/}~{\em 40}, 81--108.

\bibitem[\protect\citeauthoryear{Yu and Meng}{Yu and Meng}{2011}]{YuMen11}
Yu, Y. and X.-L. Meng (2011).
\newblock {To Center or Not to Center: That Is Not the Question---An
  Ancillarity---Sufficiency Interweaving Strategy (ASIS) for Boosting MCMC
  Efficiency}.
\newblock {\em Journal of Computational and Graphical Statistics\/}~{\em 20},
  531--570.

\end{thebibliography}

\appendix

\section{MCMC algorithms}
\label{MCMC_full}

The model is fitted using the latent variable representation of the SoftBART-CORM model in 
\eqref{def-BART-CORM-2}. The step for updating the parameters in the sampler are listed below.

\subsubsection*{Updating $\delta_k(\boldsymbol{x})$ and $\mu_k$}

The likelihood contribution of the full conditional of $\delta_k(\boldsymbol{x})$ for $i$ such that $q_i = k$ is
\[
y_i \sim  \mbox{N}\left(y_i\left\vert \mu_k + \delta_k(\boldsymbol{x}_i), \sigma^2_k\,\exp\{\phi_k(\boldsymbol{x}_i)\}\right.\right).
\]
This is a SoftBART model with observation-specific variance and the update of  $\delta_k(\boldsymbol{x}_i)$  is standard.
The full conditional distribution of $\mu_k$ is
\[
\mu_k \sim \mbox{N}\left(
\frac{\sum_{i\vert q_i = k}(y_i - \delta_k(\boldsymbol{x}_i)) / (\sigma^2_k\,\phi_k(\boldsymbol{x}_i))}{
\sum_{i\vert q_i = k} 1 / (\sigma^2_k\,\phi_k(\boldsymbol{x}_i))},
\frac{1}{
\sum_{i\vert q_i = k} 1 / (\sigma^2_k\,\phi_k(\boldsymbol{x}_i))}
\right)
\]

\subsubsection*{Updating $\phi_k(\boldsymbol{x})$ and $\sigma^2_k$}

The likelihood contribution of the full conditionals of $\phi_k(\boldsymbol{x})$  for $i$ such that $q_i = k$ is
\[
\log \left[\left(y_i - \mu_k - \delta_k(\boldsymbol{x}_i)\right)^2\right] - \log \sigma^2_k - \phi_k(\boldsymbol{x}_i) \sim \log\chi^2_1
\]
where $\log\chi^2_1$ represents the $\log \chi^2_1$ distribution ({\it i.e.} the distribution of the logarithm of a $\chi^2_1$ random variables). We use the idea of \citet{OCSN07} \citep[see also][]{KSC98} to express the $\log \chi^2_1$ distribution as a 10 component mixture of normal distributions which can be placed in a Metropolis-Hastings sampler to exactly sample from the full conditional \citep{SakGri17}. To update $\phi_k(\boldsymbol{x})$, we sample an indicator variable $t_i$, $i = 1,\dots, n$ where probability that $t_i =j$ is proportional to
\[
 w_j \,\mbox{N}\left(\left.\left(y_i - \mu_{q_i} - \delta_{q_i}(\boldsymbol{x}_i)\right)^2 + c\right\vert   \log\sigma^2_{q_i} + \phi_{q_i}(\boldsymbol{x}_i) + m_j,  v^2_j\right), \qquad j = 1,\dots, 10
\]
where $w_1, \dots, w_{10}$, $m_1, \dots, m_{10}$ and $v^2_1, \dots, v^2_{10}$ are given in \citet{OCSN07} and $c$ is a small constant chosen to avoid taking the logarithm of a number very close to zero. Conditional on $t_1, \dots, t_n$, the likelihood contribution for $\phi_k(\boldsymbol{x})$ of the $i$-th observation, if $q_i = k$, is 
\[
\log \left[\left(y_i - \mu_k - \delta_k(\boldsymbol{x}_i)\right)^2 + c\right] - \log\sigma^2_k - m_{t_i}\sim \mbox{N}\left( \phi_k(\boldsymbol{x}_i), v^2_{t_i}\right)
\]
which is again the likelihood of a SoftBART model with a observation-specific variance and is a standard update.
We update $\sigma^2_k$ using 
\[
\sigma_k^2\sim\mbox{IG}\left(a + \frac{1}{2} \sum_{i=1}^n\mbox{I}(q_i = k),
b + \frac{1}{2}\sum_{i=1}^n \mbox{I}(q_i = k)\exp\{-\phi_k(\boldsymbol{x}_i)\} \,(y_i - \mu_k - \delta_k(\boldsymbol{x}_i))^2
\right).
\]

\subsubsection*{Updating the allocation variables}

The full conditional of $q_i$ is discrete with probabilities, for $k = 1,\dots, K$, 
\[
p(q_i = k) \propto
  J_k\, \exp\{\lambda_k(\boldsymbol{x}_i)\}\,
 \mbox{N}\left(y_i\left\vert \mu_k + \delta_k(\boldsymbol{x}_i), \sigma^2_k\,\exp\{\phi_k(\boldsymbol{x}_i)\}\right.\right).
\]

\subsubsection*{Updating the weight parameters}

Firstly, to update $J_1,\dots, J_K$, we introduce latent variables $v_1, \dots, v_n$ where
\[
v_i \sim\mbox{Ex}\left(\sum_{k=1}^{K}  J_r\, \exp\{\lambda_r(\boldsymbol{x}_i)\}\right), \qquad i = 1, \dots, n
\]
and $\mbox{Ex}(\lambda)$ represents an exponential distribution with mean $1/\lambda$. The full conditional of $J_k$ is 
\[
J_k\sim\mbox{Ga}\left(\frac{M}{K} + \sum_{i=1}^n \mbox{I}(q_i = k), 1 + \sum_{i=1}^n v_i \exp\{\lambda_k(\boldsymbol{x}_i)\}\right).
\]
The likelihood contribution to the full conditional of $\lambda_k(\boldsymbol{x})$ is 
\[
\prod_{i=1}^n
\frac{\exp\{\lambda_{k}(\boldsymbol{x}_i)\}^{\mbox{I}(q_i = k)} }
{  \exp\{\lambda_k(\boldsymbol{x}_i)\} \, J_k + \sum_{m=1; m\neq k}^K \exp\{\lambda_m(\boldsymbol{x}_i)\} \, J_m}
\propto
 \prod_{i=1}^n
\frac{\exp\{\lambda_{k}(\boldsymbol{x}_i) + \log b - \log a_i\}^{\mbox{I}(q_i = k)}}
{ \exp\{\lambda_k(\boldsymbol{x}_i) + \log b - \log a_i \} + 1},
\]
where $a_i =  \sum_{m=1; m\neq k}^K \exp\{\lambda_m(\boldsymbol{x}_i)\} \, J_m$ and $b =   J_k$, 
which is a logistic regression with response $\mbox{I}(s_i = k)$, offset $\log b - \log a_i$ and 
effect
 $\lambda_k(\boldsymbol{x}_i)$. This can be update using P\'oyla-gamma data augmentation \citep{PSW13}.



\subsubsection*{Updating $\boldsymbol{s}^{\mu}$, $\boldsymbol{s}^{\phi}$ and $\boldsymbol{s}^{\lambda}$}

The full conditional distribution of $\boldsymbol{s}^{\mu}$, $\boldsymbol{s}^{\phi}$ and $\boldsymbol{s}^{\lambda}$ have the same form. Suppose that $\zeta\in\{\mu, \phi, \lambda\}$ then the full conditional of $\boldsymbol{s}^{\zeta}$  is
\[
\mbox{Dir}_p\left(a^{\zeta}_1, \dots, a^{\zeta}_p\right)
\]
where $a_k^{\zeta}  = \frac{1}{p} + n_k^{\zeta}$ and $n_k^{\zeta}$ is the number of times that variable $k$ is used in a split in a tree for $\zeta_1, \dots, \zeta_K$.

\subsubsection*{Updating $\rho^{\mu}$, $\rho^{\phi}$ and $\rho^{\lambda}$}

The full conditional distribution of $\rho^{\mu}$, $\rho^{\phi}$ and $\rho^{\lambda}$ have the same form. Suppose that $\zeta\in\{\mu, \phi, \lambda\}$ then the full conditional density of $\rho^{\zeta}$  is proportional to
\[
\left(\rho^{\zeta}\right)^{-(K+1)/2} \left[1 + 16\,R\,\rho^{\zeta}\right]^{-1}
\ \prod_{k=1}^K \left[1 + \frac{\kappa^{\zeta}_k}{\rho^{\zeta}}\right]^{-1}.
\]
This can be updated using an adaptive random walk on the logarithmic scale.

\subsubsection*{Jointly updating the offset and the level of a SoftBART}

We can write each characteristic in an additive form: $\mu_k + \delta_k(\boldsymbol{x})$, $\log \sigma^2_k + \phi_k(\boldsymbol{x})$ and $\log J_k + \lambda_k(\boldsymbol{x})$. Since each BART has prior mean zero then there is a danger that updating the offset separately from the BART can lead to poor mixing. We improve mixing by adding an extra which leaves the sum of the BART and offset unaffected but changes the level of the BART. We illustrate the approach using the mean $\mu_k + \delta_k(\boldsymbol{x})$ but it can be extended to the other characteristics.

We can introduce an extra parameter $Z$ and define $\mu_k' = \mu_k + Z_k$ and $\delta'_k(\boldsymbol{x}) = \delta_k(\boldsymbol{x}) - Z_k$. If the leaf level is $\tau_{r, l}$ for the $l$-th leaf of the $r$-th tree then we define $\tau'_{r, l} = \tau_{r, l} - Z_k / R$ which guarantees that 
$\delta'_k(\boldsymbol{x}) = \delta_k(\boldsymbol{x}) - Z$. This leaves the level of the means unchanged and the full conditional of $Z_t$ is proportional to
\[
\mbox{N}(\mu_k + Z_k\vert 0, 1) \, p_\delta(\delta_k(\boldsymbol{x}) - Z).
\]
This is available in closed form for the means and can be updated using an adaptive Metropolis-Hastings random walk on the logarithmic scale for the variances and the weights.

\subsubsection*{Updating $M$}

The full conditional density for $M$ is proportional to
\[
\frac{1}{\Gamma(M / K)^K}\left(\prod_{k=1}^K J_k\right)^{M / K} \,\exp\left\{-M \frac{1}{K}\sum_{k=1}^K J_k\right\}
\]
which can be updated using an adaptive Metropolis-Hastings random walk step on the logarithmic scale.
\section{Real data sets}

\begin{table}[h!]
\begin{tabular}{|l|l|l|}
\hline
Variable&Description&Unit of measurement\\
\hline
Normalised power output  ($y$) &$\dfrac{\mbox{actual power}}{\mbox{rated power}}\times 100$&\% of rated power\\
\hline
Wind speed ($x_1$) &meters per second &m/s\\
\hline
Wind direction ($x_2$) &measured clockwise form north& degrees\\
\hline
Air density ($x_3$) &kilograms per cubic meter & $kg/m^3$\\
\hline
Turbulence intensity ($x_4$) & $\dfrac{\mbox{S.Dev of wind speed }}{\mbox{mean wind speed}}$&  m/s\\
\hline
Below the hub shear ($x_5$) & how wind changes with hub height& dimensionless\\
\hline
\end{tabular}
\caption{Variables of Wind Turbine Data}\label{turbine}
\end{table}

\section{Additional results}

\begin{figure}[h!]
\begin{center}
\includegraphics{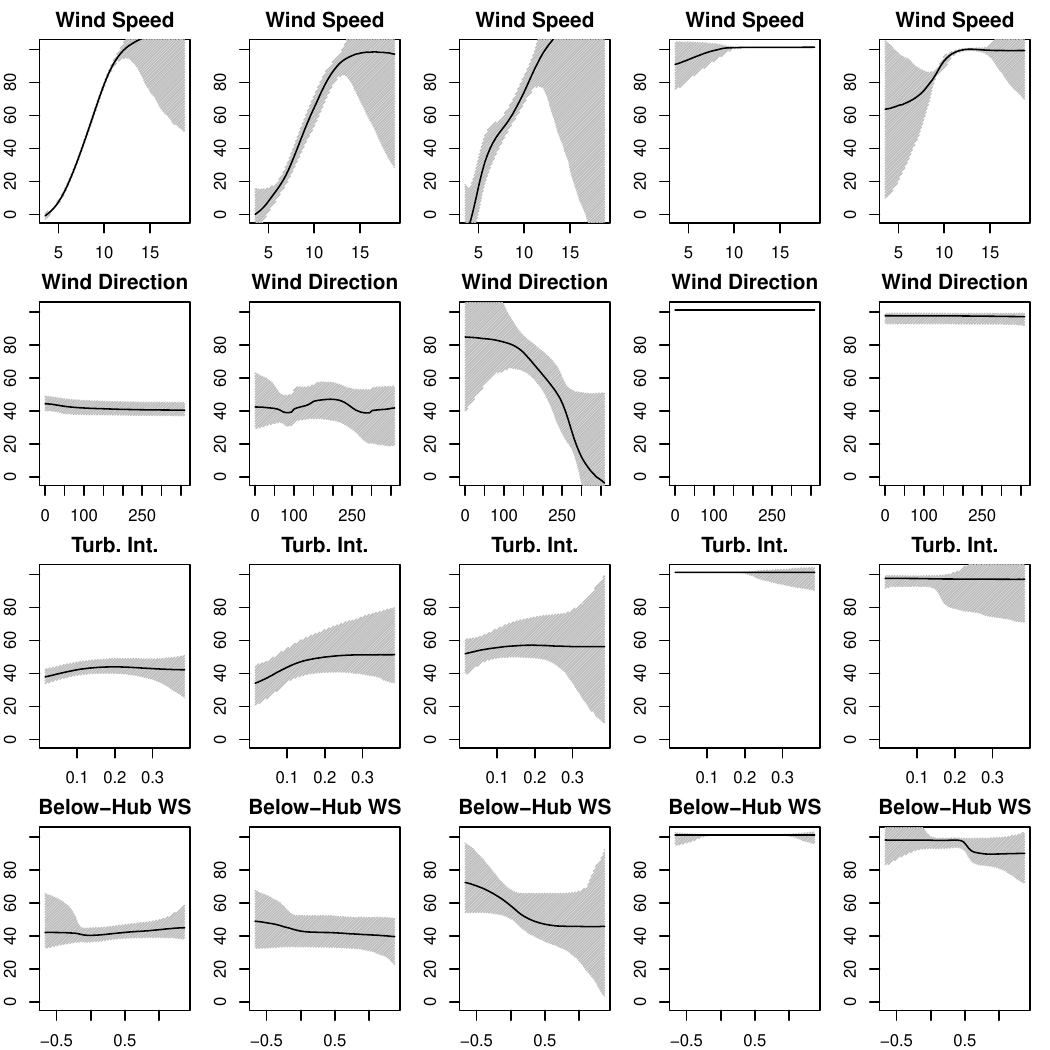}
\end{center}
\caption{Wind Turbine Prediction: The posterior median of the marginal effect of the important variables for the mean in each component with 95\% credible interval.}\label{f:wind_mean}
\end{figure}

\begin{figure}[h!]
\begin{center}
\includegraphics{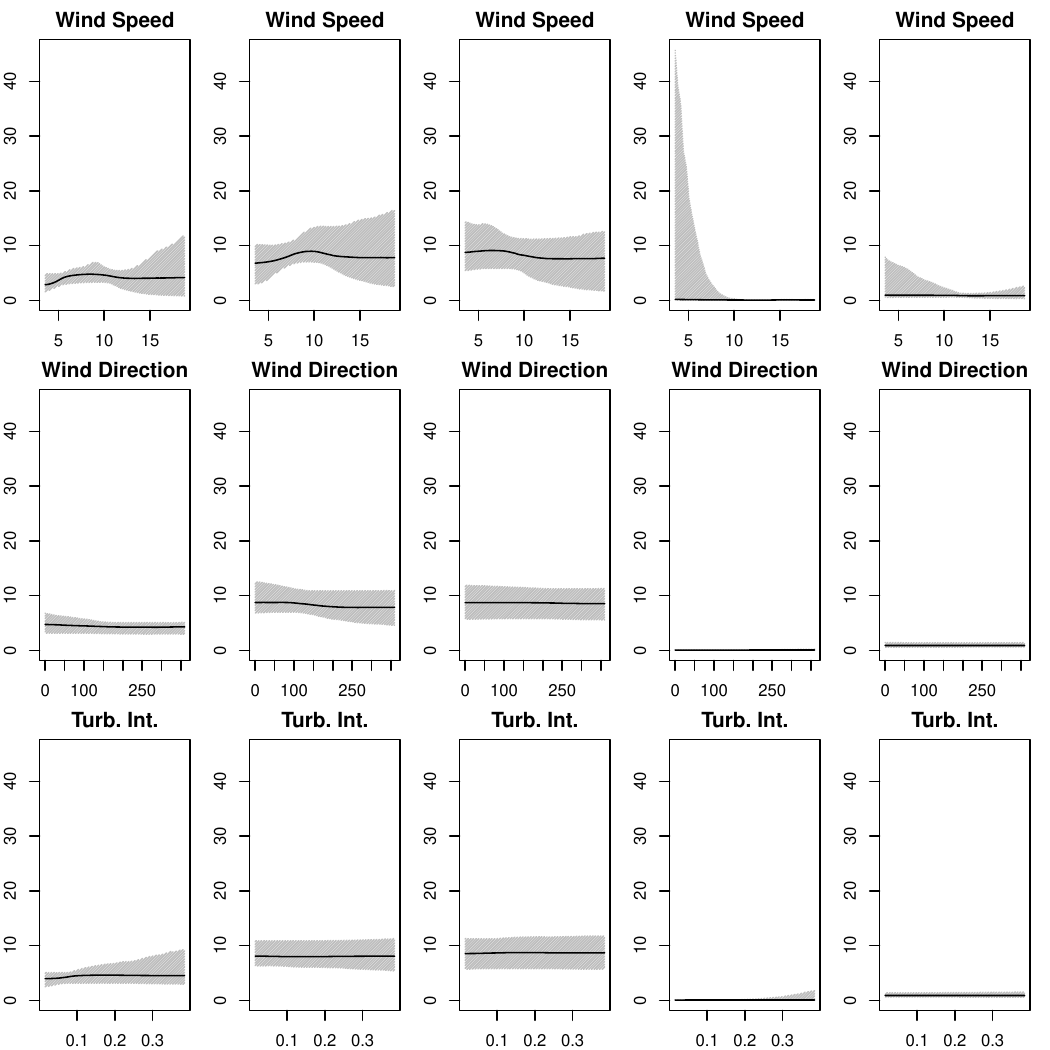}
\end{center}
\caption{Wind Turbine Prediction: The posterior median of the marginal effect of the important variables for the standard deviation in each component with 95\% credible interval.}\label{f:wind_variance}
\end{figure}

\begin{figure}[h!]
\begin{center}
\includegraphics{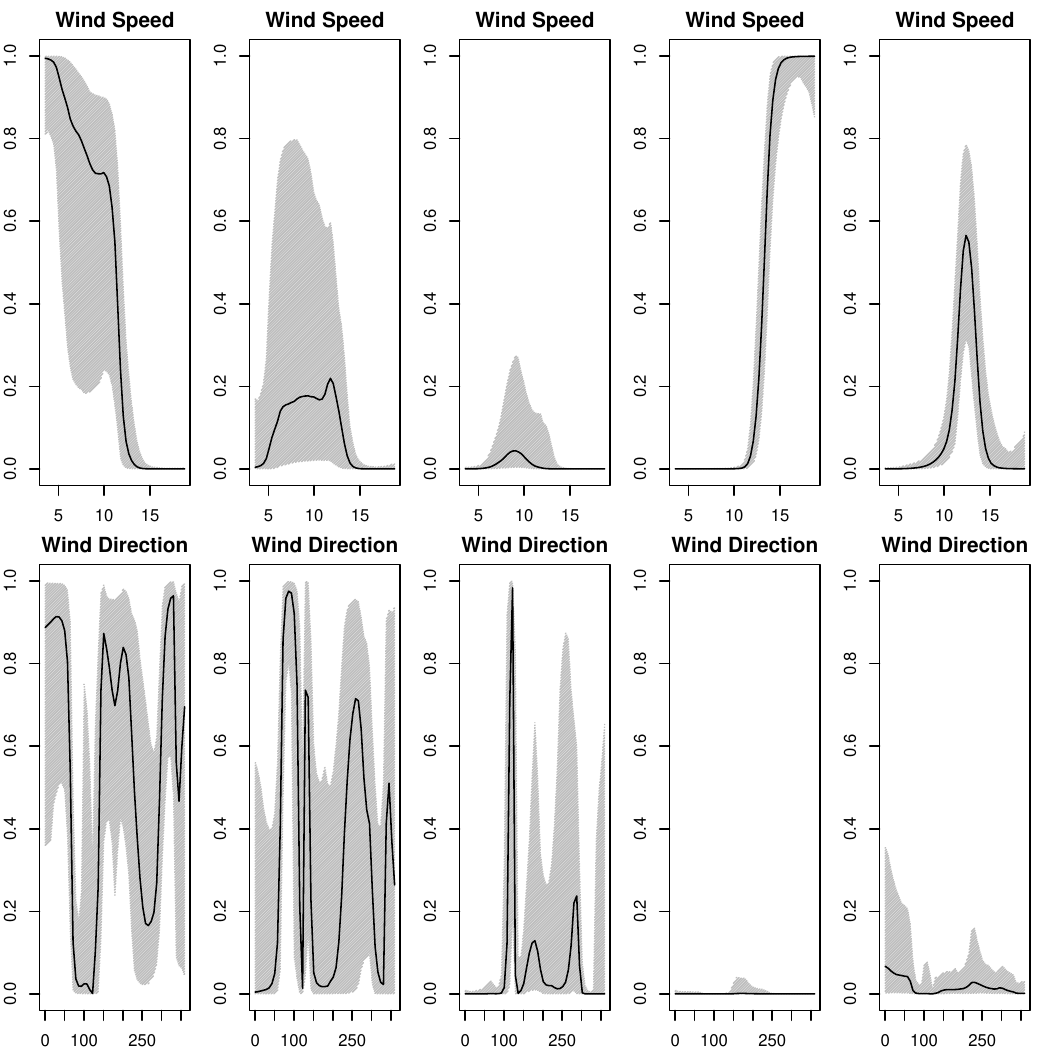}
\end{center}
\caption{Wind Turbine Prediction: The posterior median of the marginal effect of the important variables for the weights of each component with 95\% credible interval.}\label{f:wind_weights}
\end{figure}

\begin{figure}[h!]
\begin{center}
\includegraphics{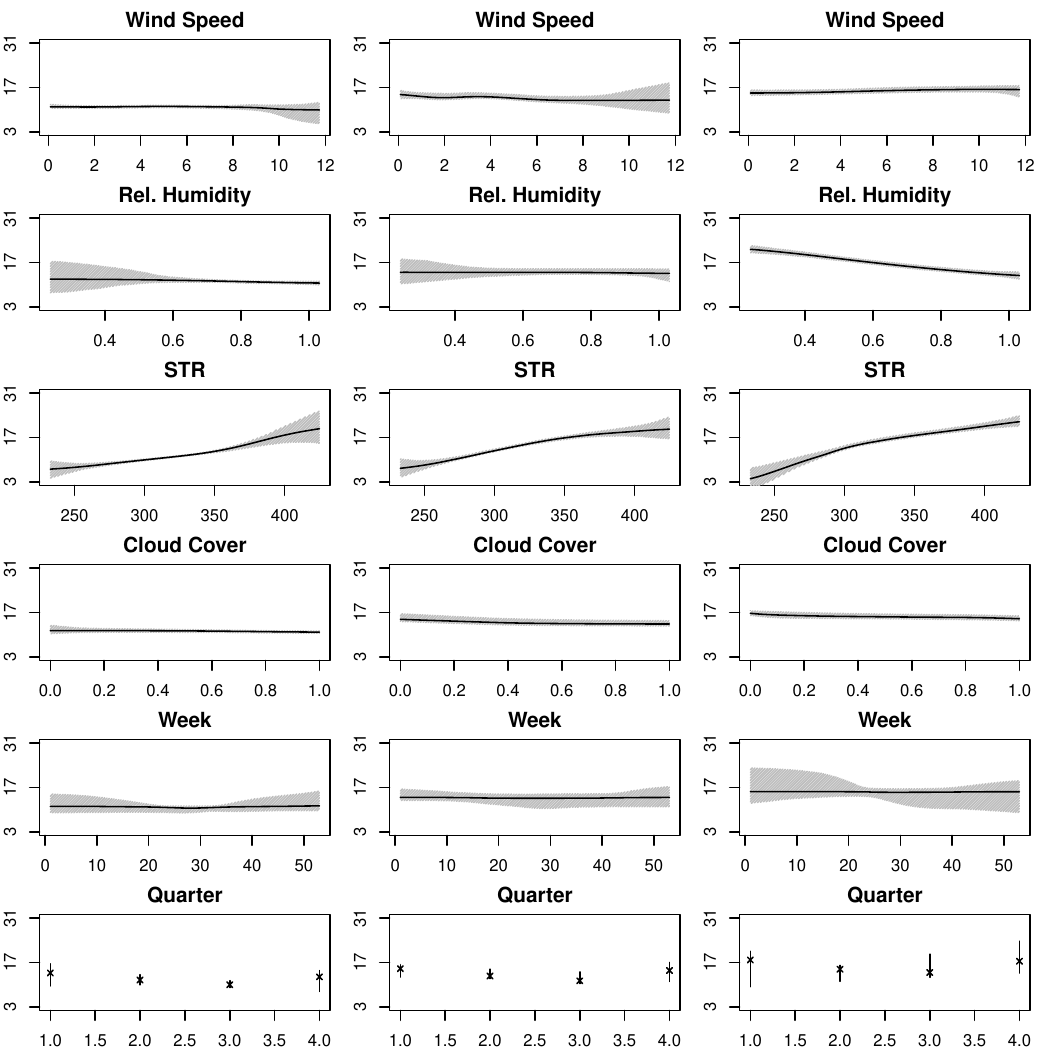}
\end{center}
\caption{Temperature Prediction: The posterior median of the marginal effect of the important variables for the mean in each component with 95\% credible interval.}\label{f:weather_mean}
\end{figure}

\begin{figure}[h!]
\begin{center}
\includegraphics{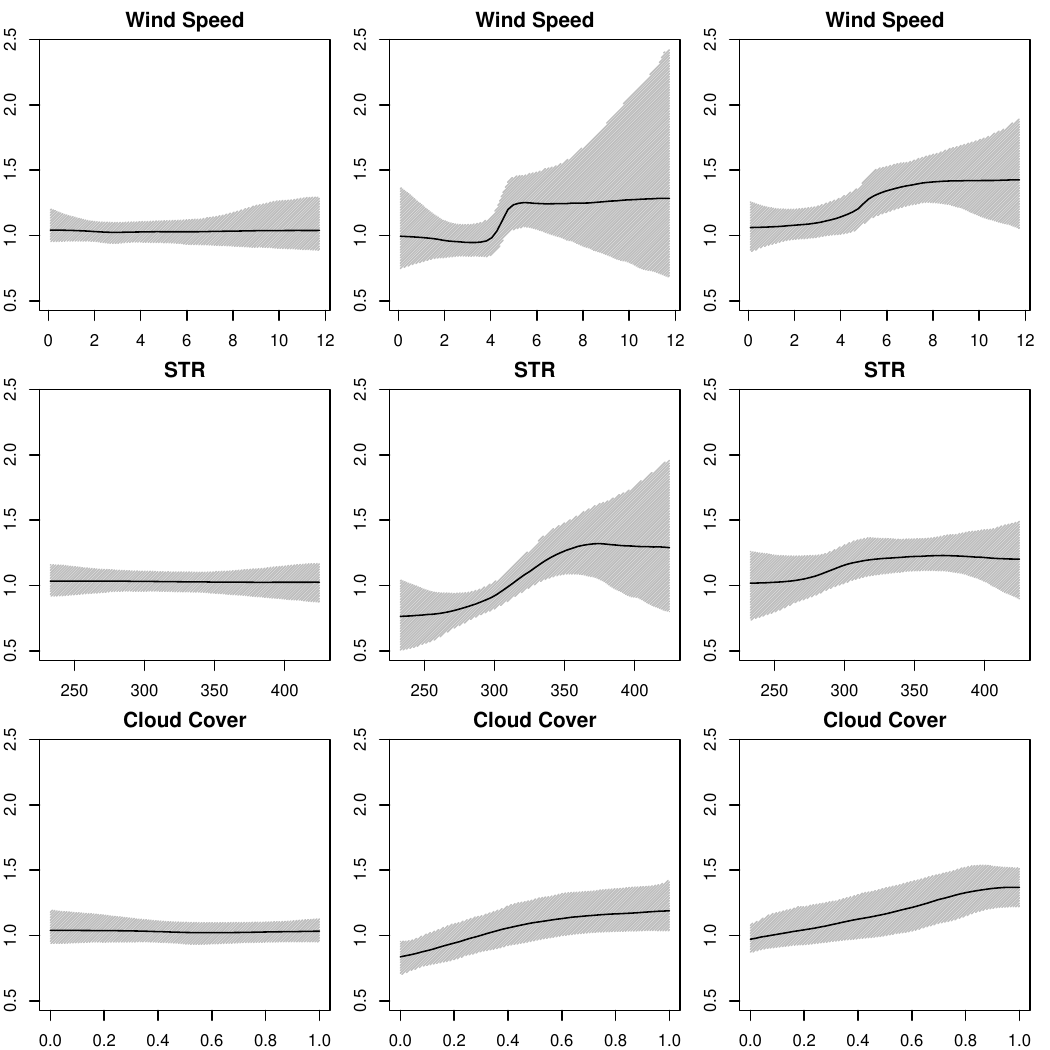}
\end{center}
\caption{Temperature Prediction: The posterior median of the marginal effect of the important variables for the standard deviation in each component with 95\% credible interval.}\label{f:weather_variance}
\end{figure}

\begin{figure}[h!]
\begin{center}
\includegraphics{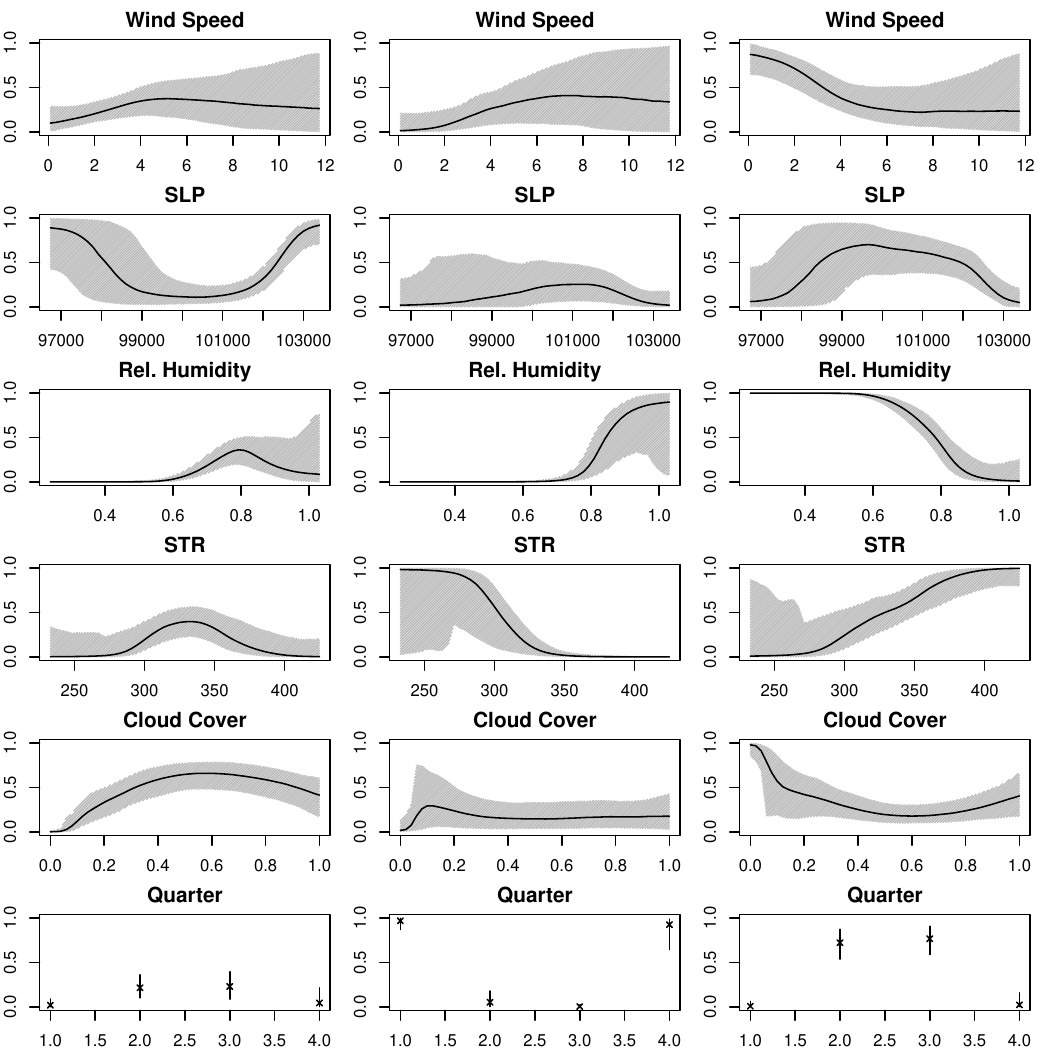}
\end{center}
\caption{Temperature Prediction: The posterior median of the marginal effect of the important variables for the weights of each component with 95\% credible interval.}\label{f:weather_weights}
\end{figure}

\end{document}